\documentclass[11pt]{article}
\usepackage[utf8]{inputenc}
\usepackage{array}
\usepackage{amsmath,amsthm,amssymb,hyperref,graphicx,physics,caption,subcaption,authblk}%, autoref}

\usepackage[capitalize]{cleveref}
\usepackage[moderate]{savetrees}
\usepackage{geometry}
\newgeometry{
    top=1.00in,
    bottom=1.00in,
    left=1.00in,
    right=1.00in
}
\usepackage{algorithm,tabularx}
\usepackage[noend]{algpseudocode}
\makeatletter
\newcommand{\multiline}[1]{%
  \begin{tabularx}{\dimexpr\linewidth-\ALG@thistlm}[t]{@{}X@{}}
    #1
  \end{tabularx}
}
\makeatother
\usepackage{setspace}
\usepackage{mathtools}
\newtheorem{theorem}{Theorem}
\newtheorem{corollary}[theorem]{Corollary}

\newtheorem{definition}[theorem]{Definition}
\newtheorem{claim}[theorem]{Claim}

\begin{document}

\title{Quantum Boosting using Domain-Partitioning Hypotheses}
\author[1]{Debajyoti Bera\thanks{dbera@iiitd.ac.in}
}
\author[2]{Rohan Bhatia\thanks{rohanbhatia\_2k19ep.079@dtu.ac.in}
}
\author[2]{Parmeet Singh Chani\thanks{parmeetsinghchani\_2k19ep.066@dtu.ac.in}
}
\author[1]{Sagnik Chatterjee\thanks{sagnikc@iiitd.ac.in}
}
\affil[1]{Indraprastha Institute of Information Technology (IIIT-D), Delhi, India}
\affil[2]{Delhi Technical University, Delhi, India}

\maketitle
\begin{abstract}
Boosting is an ensemble learning method that converts a weak learner into a strong learner in the PAC learning framework. Freund and Schapire designed the Godel prize-winning algorithm named AdaBoost that can boost learners which output binary hypotheses. Recently, Arunachalam and Maity presented the first quantum boosting algorithm with similar theoretical guarantees. Their algorithm, which we refer to as QAdaBoost henceforth, is a quantum adaptation of AdaBoost, and only works for the binary hypothesis case. QAdaBoost is quadratically faster than AdaBoost in terms of the VC-dimension of the hypothesis class of the weak learner but polynomially worse in the bias of the weak learner.

Izdebski et al.\ posed an open question on whether we can boost quantum weak learners that output non-binary hypothesis. In this work, we address this open question by developing the QRealBoost algorithm which was motivated by the classical RealBoost algorithm. The main technical challenge was to provide provable guarantees for convergence, generalization bounds, and quantum speedup, given that quantum subroutines are noisy and probabilistic. We prove that QRealBoost retains the quadratic speedup of QAdaBoost over AdaBoost and further achieves a polynomial speedup over QAdaBoost in terms of both the bias of the learner and the time taken by the learner to learn the target concept class.

Finally, we perform empirical evaluations on QRealBoost and report encouraging observations on quantum simulators by benchmarking the convergence performance of QRealBoost against QAdaBoost, AdaBoost, and RealBoost on a subset of the MNIST dataset and Breast Cancer Wisconsin dataset.

\end{abstract}

\section{Introduction}
\label{sec:Introduction}
The last decade has seen substantial growth in the field of quantum machine learning, giving rise to several quantum machine learning algorithms that promise and provide improvements over their classical counterparts. Several survey papers and books have already been published, and interested readers may consult any of those to obtain an overview of the algorithmic and theoretical advances in quantum machine learning~\cite{arunachalam2017guest,biamonte2017quantum,schuld2015introduction,wittek2014quantum}.

% \marginpar{Would boosting be helpful in a NISQ setup?}However, owing to the current state of noisy intermediate-scale quantum computers, one of the main criticisms with the field is the practical feasibility of the proposed quantum learning algorithms. A theoretically superior quantum learning algorithm may turn out to perform extremely poorly when implemented on NISQ setups, which leads us to an important question - is there a way to increase the predictive power of weak and inaccurate quantum learning algorithms? 

In the discriminative models, learning algorithms aim to ``learn" an unknown concept which helps them classify samples. Some learning algorithms are accurate with arbitrarily high accuracy, while others perform slightly better than random guessing. Even though very accurate learners are ultimately desired, it might not always be wise to use highly accurate learners for many reasons, such as longer running times,  overfitting, and lack of model explainability. On the other hand, many well-known simple learning algorithms are easy to create and are essentially weak learners. These include decision stumps, na\"{\i}ve Bayes over a single variable, clustering algorithms with a fixed number of clusters, are all . Ensemble learning is a method of converting ``weak" learners to ``strong" learners\footnote{We define the terms weak and strong formally in \cref{sec:paclearning}.}. 
%Classically, there are many ways of improving the accuracy of weak learners, such as ensemble learning methods that combine multiple weak learning algorithms to obtain a strong and accurate learning algorithm. 
% A simple ensemble learning method involves training multiple weak learners and taking a majority vote of their predictions to serve as the final prediction. We can modify this simple method to work with various forms of data by simply changing the definition of majority. A more theoretically sophisticated ensemble learning framework is \textbf{Bagging}\cite{breiman1996bagging}, where we train different weak learners on different subsets of the training set, thereby reducing variance in the final prediction. 
In this paper, we focus on a particular type of ensemble learning approach known as \textbf{Boosting}, in which weak learners are trained iteratively over reweighted distributions of a fixed training set. By tweaking the distribution of the training set, we ensure that each learner gives more weight to the misclassified samples in the previous iteration, eventually reducing both bias and variance of the learner~\cite{Friedman2000}. Boosting algorithms are now included in standard machine learning libraries, e.g., {\tt scikit-learn}~\cite{pedregosa2011scikit}.

\subsection{Related Work}
\begin{figure}
    \centering
    \includegraphics[width = 0.64\textwidth]{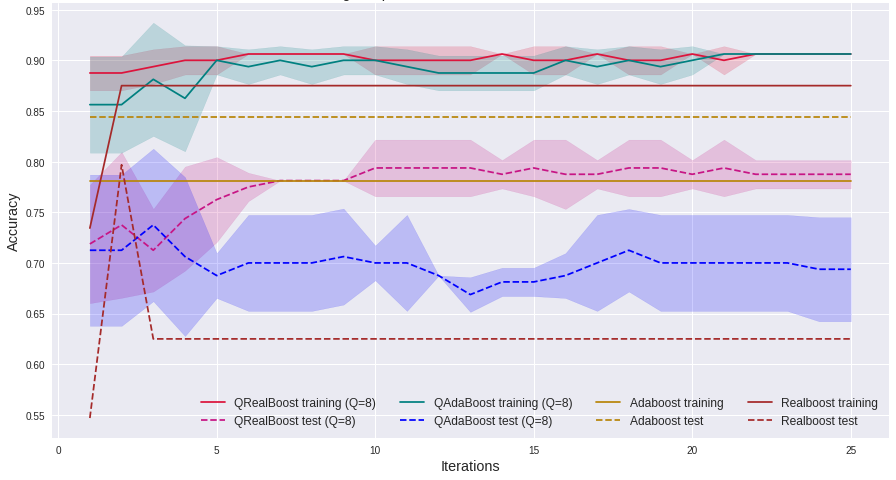}
    \caption{{\footnotesize Comparing the performance of 4 different boosting algorithms using the $k$-means clustering algorithm (with $k=3$) as the base learner on the Breast Cancer Wisconsin dataset~\cite{BreastCancer2019} with 32 training samples.}}
    \label{fig:breastcancer}
\end{figure}
\begin{table}[h!]
\small
\centering
\caption{\small 
% We compare the {query complexities} of the original AdaBoost algorithm, classical SmoothBoost algorithm, the two existing quantum boosting algorithms \cite{Arunachalam2020,deWolf2020}, and the QRealBoost algorithm that we present in this paper. Suppose we have a weak PAC learner $A$ (classical or quantum) that we are trying to boost, which has an associated hypothesis class $\mathcal{H}$. The query complexities are given in terms of the VC-dimension of $\mathcal{H}$, the bias $\gamma$ of the weak learner $A$, and the time taken (denoted by $R$ for the classical case, and $Q$ for the quantum case) by the weak learner $A$ to produce a hypothesis $h\in \mathcal{H}$.
Comparing AdaBoost, SmoothBoost, QAdaBoost, and QSmoothBoost, with the QRealBoost algorithm. Here, robust means robust to classification noise; fast and slow refer to convergence rates; binary and non-binary refer to the type hypotheses produced by the weak learner. The weak learner (classical or quantum) has bias $\gamma$, an associated hypothesis class $\mathcal{H}$ with VC-dimension of $d_\mathcal{H}$, and takes time $R$ (classical case), or $Q$ (quantum case) to produce a hypothesis $h\in \mathcal{H}$. 
}
% {\renewcommand{\arraystretch}{2}
% \begin{tabular}{p{4cm} p{3cm} p{8cm}} 
%  \hline
%  \textbf{Boosting Algorithms} & \textbf{Query Complexity} &\textbf{Remarks}\\[0.5ex] 
%  \hline
%  Classical AdaBoost (\cite{Freund1997}) &  $O\left(\frac{{d_{\mathcal{H}}}\cdot R}{\gamma^{4}}\right)$ & adaptive, very fast convergence\\
% %  [2ex]
% \hline
%  Classical SmoothBoost (\cite{servedio2003smooth}) &  $\Tilde{O}\left(\frac{{d_{\mathcal{H}}}}{\gamma^{4}}+\frac{R}{\gamma^{2}}\right)$ & robust against noise, non-adaptive, slow convergence\\
% %  [2ex]
% \hline
%  QAdaBoost (\cite{Arunachalam2020}) &  $\Tilde{O}\left(\frac{\sqrt{d_{\mathcal{H}}}\cdot Q^{1.5}}{\gamma^{11}}\right)$ & adaptive\\
% %  [2ex]
%  \hline
%  QSmoothBoost (\cite{deWolf2020})&  $\Tilde{O}\left(\frac{\sqrt{d_{\mathcal{H}}}}{\gamma^{5}}+\frac{Q}{\gamma^{4}}\right)$& robust against noise, non-adaptive\\
% %  [2ex]
%  \hline
%  QRealBoost (This paper) &  $\Tilde{O}\left(\frac{\sqrt{d_{\mathcal{H}}}\cdot Q}{\gamma^{9}}\right)$& adaptive, non-binary hypotheses\\ 
% %  [2ex] 
%  \hline
% \end{tabular}
% }
{\renewcommand{\arraystretch}{2}
\begin{tabular}{p{4cm} p{4cm} p{5.5cm}} 
 \hline
 \textbf{Boosting Algorithms} & \textbf{Query Complexity} &\textbf{Remarks}\\[0.5ex] 
 \hline
 AdaBoost (\cite{Freund1997}) &  $O\left({d_{\mathcal{H}}}\cdot R\cdot\frac{1}{\gamma^{4}}\right)$ & adaptive, fast, binary\\
%  [2ex]
\hline
 SmoothBoost (\cite{servedio2003smooth}) &  $\Tilde{O}\left(\frac{{d_{\mathcal{H}}}}{\gamma^{4}}+\frac{R}{\gamma^{2}}\right)$ & non-adaptive, robust, slow, binary\\
%  [2ex]
\hline
 QAdaBoost (\cite{Arunachalam2020}) &  $\Tilde{O}\left(\sqrt{d_{\mathcal{H}}}\cdot Q^{1.5}\cdot\frac{1}{\gamma^{11}}\right)$ & adaptive, fast, binary\\
%  [2ex]
 \hline
 QSmoothBoost (\cite{deWolf2020})&  $\Tilde{O}\left(\frac{\sqrt{d_{\mathcal{H}}}}{\gamma^{5}}+\frac{Q}{\gamma^{4}}\right)$& non-adaptive, robust, slow, binary\\
%  [2ex]
 \hline
 QRealBoost (this work) &  $\Tilde{O}\left(\sqrt{d_{\mathcal{H}}}\cdot Q\cdot\frac{1}{\gamma^{9}}\right)$& adaptive, fast, non-binary\\ 
%  [2ex] 
 \hline
\end{tabular}
}
\label{table:querycomplexity}
\end{table}
{AdaBoost is one of the first adaptive boosting algorithms, and was proposed by Freund and Schapire \cite{Freund1997}. AdaBoost provably converges with zero training error and requires no prior knowledge about the accuracy of the hypotheses generated by the weak-learner it is trying to boost. It has been also observed that AdaBoost does not tend to overfit\cite{bauer1999empirical,drucker1996boosting} on a wide variety of problems and performed much better\cite{dietterich2000experimental,opitz1999popular} than other ensemble methods in the absence of classification noise. We discuss in more detail the original AdaBoost algorithm in \cref{sec:adaboost_gen}.}
% \marginpar{Move AdaBoost etc. to Appendix} 
Early works on quantum boosting algorithms consider quantum algorithms for AdaBoost and its variants in the experimental setting \cite{pmlr-v25-neven12}, use classical AdaBoost as a subroutine \cite{Schuld2018}, or only consider speedups for a particular aspect of AdaBoost, e.g., computing the margins as defined in \cref{alg:AdaBoost}~\cite{Wang2020}. Arunachalam and Maity recently adapted the original AdaBoost algorithm into QAdaBoost\cite{Arunachalam2020}  and provided rigorous mathematical guarantees of speed up over the classical version in its query and time complexity in terms of the weak learner’s sample complexity by using approximate counting of quantum states. Subsequently, Izdebski et al.~\cite{deWolf2020} proposed a quantum variant of Servedio's classical SmoothBoost algorithm\cite{servedio2003smooth} (which we refer to as QSmoothBoost) which retains the speedup in sample complexity while also achieving a polynomial speedup in the bias of the weak learner as compared to the QAdaBoost algorithm. Despite QSmoothBoost's impressive speedups in complexity, it comes with a few shortcomings. Similar to its classical counterpart, QSmoothBoost is not adaptive\footnote{For a detailed discussion on ``adaptiveness'' see \cite{bshouty2002boosting}}. Additionally, SmoothBoost (hence, QSmoothBoost) takes longer to converge than AdaBoost (hence, QAdaBoost) and does not converge to zero training error. We chose to stick to the AdaBoost framework to circumvent these issues\footnote{We describe a generalized framework in \cref{algo:adaboost_framework} in \cref{sec:adaboost_gen}}. In \cref{table:querycomplexity}, we compare the query complexities of our quantum boosting algorithm with the QAdaBoost, QSmoothBoost, and the classical AdaBoost algorithms.

% it retains a few disadvantages from its classical source. SmoothBoost is not fully adaptive since it requires the algorithm to be supplied with two parameters -- one parameter for bounding the margin of the final hypothesis and another for controlling the empirical error. Additionally, the SmoothBoost algorithm takes longer to converge than the AdaBoost algorithm and its variants\cite{meir2003introduction}. Furthermore, SmoothBoost by design does not ensure zero training error.  
%The novel contribution of this work is a quantum algorithm for boosting weak learners whose hypotheses essentially outputs a partition of the domain (also known as ``domain-partitioning hypotheses); such hypotheses enable confidence-rated predictions\footnote{We discuss the notion of ``confidence'' and ``confidence-rated predictions'' in  \cref{sec:BoostinDomainPartition}.}.

% A few quantum boosting algorithms predate the QAdaBoost algorithm. These  

\subsection{Our work}
%------------------------
%------------------------
\algdef{SN}{PreLoop}{EndPreLoop}[1]{\textbf{On the} \(\mbox{#1}\) \textbf{do}}{}%
\makeatletter
\algnewcommand{\LineComment}[1]{\Statex \hskip\ALG@thistlm \(\triangleright\) #1}
\makeatother
\begin{algorithm}[t]
\caption{The QRealBoost algorithm.}
\begin{algorithmic}[1]
\footnotesize
\label{alg:QRealBoost}
\Statex
\textbf{Input:}  Quantum weak PAC learner $A$ with sample complexity $Q$ which makes at most $C$ partitions in the worst case, and $M$ labeled training samples $\{(x_i,y_i)\}_{i\in[M]}$.
\State
\textbf{Initialize:}  Worst-case values of $Q$ and $C$, and $\varepsilon=O\left(\frac{1}{QT^2}\right)$,  $\kappa=\frac{C}{\left(1-\varepsilon\right)}\sqrt{\frac{1+\varepsilon}{1-\varepsilon}}$. Set $\Tilde{D}^{1}_{i}=\frac{1}{M}\;\;\forall i\in [M]$.

\For{$t=1$ to $T$}\Comment{ {\footnotesize \textbf{Adaptiveness}: Choose $T\geq\frac{\ln{M}}{2\gamma^2}$ for a worst-case guess of $\gamma$.}} 
    \Statex \quad {\bf\underline{\small generate $Q+2C$ copies of a distribution of training samples}}
    \State Prepare $\ket{\phi_0}^{\otimes Q} \ket{\psi_0}^{\otimes 2C}$ where $\ket{\phi_0}=\ket{\psi_0} = \tfrac{1}{\sqrt{M}} \sum_{i \in [M]} \ket{x_i,y_i} \otimes \ket{\Tilde{D}^1_i}$.
    % \State Prepare $Q+2C$ copies of $\phi_0=\psi_0 = \tfrac{1}{\sqrt{M}} \sum_{i \in [M]} \ket{x_i,y_i} \otimes \ket{\Tilde{D}^1_i}$.
    %\State\multiline{ Use the oracles $O_{h_1},\ldots,O_{h_{t-1}}$ and stored values of $\beta^{\prime}_{j,s}, \forall j={1,\ldots,C}$ and $b\in\{-1,+1\}$ and $Z^{\prime}_{t},\forall t\geq 1$ to create the state $\ket{\phi_2}$ using the unitary $U_D$ as follows:}
    \For{$s=1$ to $t-1$}
        \State\multiline{On all the $Q+2C$ copies, apply the transformation 
        $\ket{x_i,y_i} \ket{\Tilde{D}^{s}_i} \xrightarrow[]{} \ket{x_i,y_i} \ket{\Tilde{D}^{s+1}_i}$ 
        based on the stored values of $\beta^{\prime}_{j,s}$ and $Z^{\prime}_{s}$, obtaining the partition $j$ using $O_{h_{s}}$, and using the {\em distribution update rule} 
        \begin{equation}\label{eq:NewUpdateRule}
        % \mbox{ using the distribution update rule~}
            \Tilde{D}^{s+1}_i=\frac{\Tilde{D}^{s}_{i}\cdot\mathrm{exp}\left(-\beta^{\prime}_{j,s}y_i\right)}{\kappa Z^{\prime}_{s}}
        \end{equation}
    % \qquad using the oracle for the $(s+1)$-th hypothesis $O_{h_{s+1}}$, the stored values of $\beta^{\prime}_{j,s}$ and $Z^{\prime}_{s}$.
        }
    \EndFor
    \LineComment{The final state after the loop is denoted $\ket{\phi_2}^{\otimes Q} \ket{\psi_2}^{\otimes 2C}$ which is set to $\ket{\phi_0}^{\otimes Q} \ket{\psi_0}^{\otimes 2C}$ when $t=1$.}
    \Statex
    \PreLoop{first $Q$ copies}
        \State Perform a conditional rotation on the         $\ket{\Tilde{D}^{t}_{i}}$ register to create $\ket{\phi_3}$.
        \State Apply amplitude amplification conditioned on the $\ket{1}$ register and then uncompute ancilla to obtain $Q$ copies of $\ket{\phi_5}=\sum_{i\in [M]}\sqrt{\Tilde{D^{t}_{i}}}\ket{x_i, y_i}+\ket{\zeta_t^{\prime}}$.\label{step1}   
    \EndPreLoop
    % \State\multiline{{\bf On the first $Q$ copies:} Perform a conditional rotation on the $\ket{\Tilde{D}^{t}_{i}}$ register to create the state $\ket{\phi_3}$.}
    % \State\multiline{{\bf On the first $Q$ copies:}  Apply amplitude amplification and uncompute the ancilla registers to create the state $\ket{\phi_5}$.}\label{step1}
    % \Statex 
    \Statex {\bf\underline{\small obtain the $t$-th hypothesis $h_t$ from the first $Q$ copies}}
    \State $A \left( \ket{\phi_5}^{\otimes Q} \right) \rightarrow$ (followed by a measurement) oracle $O_{h_t}$ corresponding to the hypothesis $h_t$.\label{step2}
    % \Statex
    \Statex {\bf\underline{\small obtain confidence-rated predictions using $h_t$ on the last $2C$ copies}}
    \State\multiline{On the last $2C$ copies of $\ket{\psi_2}$ use oracles $O_{h_1},\ldots,O_{h_{t-1}},O_{h_t}$ to create $2C$ copies of $\ket{\psi_3}=\frac{1}{\sqrt{M}}\sum_{i\in [M]}\ket{x_i, y_i, \Tilde{D^{t}_{i}}}\ket{j^{1}_{i},j^{2}_{i},\ldots,j^{t}_{i}}$.}
    \For{$k=1$ to $C$ and $b\in\{-1,+1\}$}\Comment{ {\footnotesize iterate over every partition and label}}
        \State\multiline{ Take the $(k,b)$\textsuperscript{th} copy of $\ket{\psi_3}$ and prepare (omitting unrelated qubits) \\ 
            $\ket{\psi_4}=\frac{1}{\sqrt{M}}\sum_{i\in[M]}\ket{x_i}\ket{y_i}\ket{\Tilde{D}_i^{t}}\ket{j^{t}_{i}}\ket{\Tilde{D}^{k,b,t}_{i}}$ where $\Tilde{D}^{k,b,t}_{i}=\Tilde{D}_i^{t}$ if $j_i^{t}=k$ and $y_i=b$, and 0 otherwise.
            }
        \State\multiline{Do conditional rotation on the last register to obtain states with amplitudes $\sqrt{{W}^{k,t}_{b}}$ where ${W}^{k,t}_{b}={\sum_{i\in[M]}{\Tilde{D}^{k,b,t}_{i}}}/{M}$.}
        \State {Perform amplitude estimation to obtain $\Tilde{W}^{k,t}_{b}$ with relative error $\varepsilon$ and do Laplace correction on the estimated weights $\Tilde{W}^{k,t}_{b}$.}\label{step3}
    \EndFor
    \State\multiline{ Compute $Z^{\prime}_t=2\sum_{j=1}^{C}\sqrt{\Tilde{W}^{j,t}_{+}\Tilde{W}^{j,t}_{-}}$ and  margins $\beta^{\prime}_{j,t}=\frac{1}{2}\ln{\left({\Tilde{W}^{j,t}_{+}}/{\Tilde{W}^{j,t}_{-}}\right)}$ for all $j=1,\ldots,C$.}
%   \State\multiline{ Perform the distribution update $\forall i\in\{1,\ldots,M\}$ as follows
    % \begin{equation}\label{eq:NewUpdateRule}
        % \Tilde{D}^{t+1}_i=\frac{\Tilde{D}^{t}_{i}\cdot\mathrm{exp}\left(-\beta^{\prime}_{j,t}y_i\right)}{\kappa Z^{\prime}_{t}}
    % \end{equation}}
\EndFor
\Statex
\State \textbf{Output:} Hypothesis $H(\cdot)$ defined as% and $\ket{x,0}\xrightarrow[]{O_{h_t}}\ket{x,j^t}$ for all $t\in\{1,\ldots,T\}$, defined as
\begin{equation}
\label{eq:qrealboostcombinedclassifier}
    H(x)=\mathrm{sign}\left(\sum_{t=1}^{T}\beta^{\prime}_{j^t,t}\right), \mbox{~where $j^t \in \{1 \ldots C\}$ is obtained using $\ket{x,0}\xrightarrow[]{O_{h_t}}\ket{x,j^t}$}
\end{equation}
% \EndProcedure
\end{algorithmic}
\end{algorithm}
%
%---------------------
%---------------------
\label{sec:ourwork}
Our algorithm QRealBoost\cref{alg:QRealBoost} is a quantum adaptation of the discrete RealBoost algorithm~\cite{Friedman2000,Schapire1999} which tackles the problem of boosting weak learners whose hypotheses essentially divide the domain $\mathcal{X}$ into a small number of mutually exclusive and exhaustive sets of partitions (hence the name ``domain-partitioning hypotheses''). A point to note here is that such \textit{domain-partitioning} learners can be alternatively characterized as having hypotheses that output real-valued predictions, unlike binary-valued predictions as seen in AdaBoost (and QAdaBoost). 

In this work we focus on weak learners that output discrete class partitions rather than class probabilities since this is a more natural model for decision tree algorithms and clustering algorithms. Domain-partitioning hypotheses allow us to calculate the confidences of prediction for each partition~\footnote{We discuss the notion of ``confidence'' and ``confidence-rated predictions'' in  \cref{sec:BoostinDomainPartition}.} which leads to improved estimation of margins, ultimately producing better bounds of generalization error. RealBoost not only retains the zero training error and the generalization behavior of the original AdaBoost family of boosting algorithms but also has been observed in practice to converge much faster than AdaBoost with respect to the empirical error \cite{5656917,wang20092d,1301512}. QRealBoost maintains the general flavour of RealBoost but implements several steps using quantum algorithms which lead to a quadratic speedup over RealBoost. The caveat here is that the intermediate quantum subroutines involving quantum amplification and estimation are erroneous, and therefore require careful analysis to prove that boosting converges and the convergence is exponentially fast. This is the main technical contribution of this work. This is also the first work to our knowledge which performs boosting with non-binary classifiers, which also addresses an open question by Izdebski et al.~\cite{deWolf2020} on whether we can boost learners with range other than $\{-1,+1\}$. We now state the main results of our paper.
\begin{theorem}[Complexity of QRealBoost]
Let $A$ be a $\gamma$-weak quantum PAC learner for a concept class $\mathcal{C}$ with sample complexity $Q$ having an associated hypothesis class $\mathcal{H}$ with VC-dimension $d_{\mathcal{H}}$. Further, suppose that $\mathcal{H}$ contains domain-partitioning hypotheses. The time complexity of boosting $A$ to a strong PAC learner according to \cref{alg:QRealBoost} is $\Tilde{O}\left(\sqrt{d_{\mathcal{H}}}\cdot Q\cdot\frac{n^2}{\gamma^9}\right)$ and the corresponding query complexity is $\Tilde{O}\left(\sqrt{d_{\mathcal{H}}}\cdot Q\cdot\frac{1}{\gamma^9}\right)$.
\end{theorem}

\begin{theorem}[Convergence of QRealBoost]
If we run \cref{alg:QRealBoost} for a sufficiently large number of iterations $T\geq\frac{\ln{M}}{2\gamma^2}$, then with a high probability we output a hypothesis $H$ that has zero training error and a small generalization error.
\end{theorem}
\subsection{Techniques}
Our weak PAC learner makes a constant $C$ number of domain partitions. The QRealBoost algorithm focuses on improving the confidence of prediction in each partition individually. To do this we start by calculating the number of samples belonging to a given partition with a particular label. This is known as the partition-label weight. The main idea of this algorithm is to use quantum estimation techniques to calculate the $2C$ partition-label weights ${W}^{j,t}_{b}$ (as indicated in \cref{eq:realboostpartition}) for each $j\in\{1,\ldots C\}$ and $b\in\{-1,+1\}$. Depending upon the technique used for estimation, we can obtain a quadratic speedup compared to classical techniques which take $\Theta(M)$ time. Earlier quantum boosting algorithms like QAdaBoost and QSmoothBoost also employ quantum estimation in a similar manner. The challenge in those and this work is to prove that boosting takes place despite the resulting errors from the estimations. We now discuss several caveats arising from estimating the weights, which our algorithm addresses. 
\begin{itemize}
    \item \textbf{Confidence-rated predictions}: If we naively estimate the partition-label weights ${W}^{j,t}_{b}$, then there is no way to bound the estimated confidence-rated predictions $\beta^{\prime}_{j,t}$ with respect to the actual confidence-rated predictions $\beta_{j,t}$.
    In \cref{clm:marginbound} we show that we can bound the latter by carrying out relative error estimation of the partition-label weights and then performing Laplace correction to deal with corner cases where the values of ${W}^{j,t}_{b}$ might be extremely small\footnote{Note that this issue does not arise due to our algorithm, but is in fact a property of the underlying classical algorithm itself \cite{10.5555/2207821,Schapire1999}. Laplace correction is well studied in the machine learning literature, especially with respect to decision tree classifiers \cite{provost2003tree} which naturally behave as domain-partitioning learners.} (which may lead to unbounded confidence rated predictions).
    
    % The first issue we run into while estimating the partition-label weights ${W}^{j,t}_{b}$ concerns the confidence-rated predictions we make for each partition. If we simply estimate the weights naively (with additive error), i.e. $\left|{W}^{j,t}_{b}-\Tilde{W}^{j,t}_{b}\right|\leq \varepsilon$, then there is no way to bound the estimated confidence-rated predictions $\beta^{\prime}_{j,t}$ with respect to the actual confidence-rated predictions $\beta_{j,t}$. 
    
    % To address this issue, we perform a relative error estimation of ${W}^{j,t}_{b}$ to get $\left|{W}^{j,t}_{b}-\Tilde{W}^{j,t}_{b}\right|\leq \varepsilon {W}^{j,t}_{b}$. This allows us to bound the magnitude of the estimated confidence-rated predictions $\left|{W}^{j,t}_{b}-\Tilde{W}^{j,t}_{b}\right|\leq 0.1$ when $\varepsilon\leq 0.1$. Moreover, $\varepsilon$ is initialized to a much smaller fraction in our algorithm. Therefore, according to \cref{clm:marginbound} we show that our confidence-rated predictions are bounded.
    
    % A small point to note here is that we might run into cases where the values of ${W}^{j,t}_{b}$ might be extremely small, which may lead to very large margins. Note that this issue does not arise due to our algorithm, but is in fact a property of the classical algorithm itself \cite{10.5555/2207821,Schapire1999}. We perform Laplace correction of the partition weights before using them for calculating the confidence-rated predictions $\beta^{\prime}_{j,t}$.
    
    \item \textbf{Bounding the weights for each iteration}: A similar issue arises for the distribution weights for the next iteration, if we try to naively estimate the partition-label weights ${W}^{j,t}_{b}$. We can only additively bound the new normalization constant $Z^{\prime}_t$ with respect to $Z_t$ which does not guarantee that the updated weights $\Tilde{D}^{t+1}_{i}$ (for $t\geq 1$) are normalized or even sub-normalized. In \cref{clm:normconstbound} we show that relative error estimation of ${W}^{j,t}_{b}$ can bound the quantity $Z^{\prime}_t$ with relative error. We define the new distribution update rule \cref{eq:NewUpdateRule}, using which we prove \cref{clm:bound_on_subnormalized_sum} which states that the sum of weights in the next iteration is bounded in the range $\left[1-\frac{4\varepsilon}{1+\epsilon},1\right]$; this, in turn, guarantees that the resulting weights are close to a distribution.
    
    \item\textbf{Behaviour of intermediate hypotheses}: Assume that the quantum weak-learner $A$ outputs a hypothesis $h_t$ with high probability when given the ``ideal'' quantum state in which the probability (representing the weight) of each sample comes exactly from the distribution computed by RealBoost. We show in \cref{clm:nonidealhypothesis} that even when a ``non-ideal'' state, in which the probabilities are estimated with relative error, is passed by \cref{alg:QRealBoost} to the weak learner, it still outputs the same hypothesis $h_t$ with high probability.
    
    \item\textbf{Final hypothesis is good}: Even with \cref{clm:nonidealhypothesis}, we still have to prove that our combined classifier $H$ satisfies an arbitrarily high number of training samples since the base classifiers themselves are weak. Using \cref{clm:zerotrainingerror}, we show that our combined classifier has a very small generalization error.
\end{itemize}

\section{Preliminaries}
\label{sec:Preliminaries}
\subsection{PAC Learning}
\label{sec:paclearning}
% A concept $c:\mathcal{X}_n\xrightarrow[]{}\{-1,+1\}$ is an $n$-bit Boolean function over some domain $\mathcal{X}$. A concept class $\mathcal{C}$ is a family of concepts, where $C=\bigcup_{n\geq 1}C_n$ and $C_n$ is the set of concepts over the domain $\mathcal{X}_n$. Let $D$ be an unknown probability distribution over the set of points $\mathcal{X}=\bigcup_{n\geq 1}\mathcal{X}_n$, i.e. $D:\mathcal{X}_n\xrightarrow[]{}[0,1]$.

We assume familiarity with the PAC learning framework but we quickly go over the basic concepts. A concept class $\mathcal{C}$ is a family of concepts and each concept $c$ is a set of $n$-bit Boolean functions, one for each $n$. Suppose we are given a set of $M$ labelled examples $S=\{(x_i,y_i=c(x_i))| i\in[M]\}$ (which we call the training data) taken from an unknown distribution and an unknown concept $c$ from a concept class $\mathcal{C}$ where $x_i\sim D$. The objective of PAC learning is to ``learn'' the unknown target concept from the training data such that the ``learned'' concept generalizes well to all points sampled from the same distribution $D$. 
% We do this by choosing a learning algorithm that produces hypotheses of a certain form, i.e., from a set of different hypotheses; we denote this class of hypotheses associated with a learning algorithm $\mathcal{H}$. 
We denote a learning algorithm $A$ as an $(\eta,\delta)$-PAC learner for the concept class $\mathcal{C}$, if it efficiently\footnote{The learner runs in time polynomial in $M, 1/\eta, 1/\delta$, and $n$.} outputs a hypothesis $h$ such that with probability at least $1-\delta$ (over its internal randomness) $\displaystyle \underset{x\sim D}{\mathrm{Pr}}\left[h(x)\neq c(x)\right]\leq \eta$. 

A $\gamma$-\textbf{weak learner} $A$ is defined as a $\left(\frac{1}{2}-\gamma,\delta\right)$-PAC learner, where $\gamma=O(1/n^k)$ for $k\geq 1$ and a \textbf{strong learner} is defined as a $\left(\frac{1}{3},\delta\right)$-PAC learner, using $\delta\leq 1/3$ in both cases. One of the most celebrated results of PAC learning was produced by Schapire who showed that under the PAC learning model, the task of producing strong learners from weak learners is not only possible but that the two notions of learning are inherently equivalent~\cite{Schapire1990}.
% \begin{equation}
% \label{eq:etadeltalearner}
%   \underset{x\sim D}{\mathrm{Pr}}\left[h(x)\neq c(x)\right]\leq \eta.
% \end{equation}
% \marginpar{for all $c \in C$?}  
% Alternatively, we say that the concept class $\mathcal{C}$ is $(\eta,\delta)$-PAC learnable if there exists a learning algorithm $A$ which efficiently produces a hypothesis $h:\mathcal{X}_n\xrightarrow[]{}\{-1,+1\}$ which are $\eta$ close to $c\in C_n$ with probability $1-\delta$ for all distributions $D:\mathcal{X}_n\xrightarrow[]{}[0,1]$. 
% Since the PAC learning model does not depend on the distribution $D$ from which our examples are sampled to find the target concept class $\mathcal{C}$, we call it a distribution-free model. 
In order to quantify how accurate our hypothesis is, we define two types of misclassification errors: \textbf{training error} $\Hat{\mathrm{err}}(h,S)=\underset{(x,y)\in S}{\mathrm{Pr}}\left[h(x)\neq y\right]$ which is defined with respect to the training set, and \textbf{generalization error} $\mathrm{err}(h,D)=\underset{(x,y)\sim D}{\mathrm{Pr}}\left[h(x)\neq y\right]$ which is defined with respect to arbitrary samples.
% \begin{equation}
% \label{eq:trainerror}
%     \Hat{\mathrm{err}}(h,S)=\Hat{\mathrm{err}}(h)=\underset{(x,y)\in S}{\mathrm{Pr}}\left[h(x)\neq y\right]
% \end{equation}
% and generalization error as
% \begin{equation}
% \label{eq:testerror}
%     \mathrm{err}(h,C,D)=\mathrm{err}(h)=\underset{(x,y)\sim D}{\mathrm{Pr}}\left[h(x)\neq y\right].
% \end{equation}

% Before we move forward, let us formally define the notion of weak learners and strong learners.
% \begin{definition}[$\gamma$-weak learner]
% A $\gamma$-weak learner $A$ having an associated hypothesis class $\mathcal{H}$ for a target concept class $\mathcal{C}=\bigcup_{n\geq 1}C_n, n\geq 1$ is a $\left(\frac{1}{2}-\gamma,\delta\right)$-PAC learner, where $\gamma=O(1/n^k), k\geq 1$. We assume that $\delta\leq 1/3$ in this case.
% \end{definition}
% \begin{definition}[Strong learner]
% A strong learner $A$ having an associated hypothesis class $\mathcal{H}$ for a target concept class $\mathcal{C}=\bigcup_{n\geq 1}C_n, n\geq 1$ is a $\left(\frac{1}{3},\delta\right)$-PAC learner. As before, we assume that $\delta\leq 1/3$ in this case.
% \end{definition}

\subsubsection{Quantum PAC learning}
\label{sec:quantumpaclearning}
Our algorithm is designed in the quantum PAC learning framework, introduced by Bshouty et al.~\cite{bshouty1998learning}, but now extended to learners using quantum examples. In the classical setting a learner $A$ can query multiple samples from $S$ (we denote its sample complexity as $Q$), while in the quantum setting~\cite{Arunachalam2020,deWolf2020} we assume that the examples are provided in the form of the state $\left( \tfrac{1}{\sqrt{M}} \sum_{x_i \in S} \ket{x_i, y_i=c(x_i)} \right)^{\otimes Q}$. We observe that in order to simulate a classical learner, a quantum learner can measure this state to obtain $Q$ examples chosen uniformly at random (with replacement) from $S$.
This state can be efficiently prepared with or without the assumption of a quantum random access memory ({\it aka.} QRAM). To use a QRAM to prepare a uniform superposition over the classical samples, we only incur an additive $O(\sqrt{M}\log M)$ term in the query complexity which retains our quantum speedup. We also note that the QRAM (if used) is only for state preparation. 
% and not as an oracle for search problems\footnote{This bypasses some issues observed by Arunachalam et al.~\cite{arunachalam2015robustness}.}. 
%
For a detailed discussion on the preparation of quantum samples without a QRAM, we refer the reader to Izdebski et al.~\cite{deWolf2020}, in which the authors only assume quantum query access to the training samples.

% A natural extension to the $M$ training samples in the classical setting from which $A$ can sample
% Now, we briefly discuss the Quantum PAC learning model. While in the classical setting we had the set of $M$ training samples $S$, in the quantum setting we assume we have access to quantum examples with respect to some distribution $D$ as 
% \begin{equation*}
%     \sum_{x\in \mathcal{X}} \sqrt{D_x} \ket{x,c(x)}
% \end{equation*}
% where $c\in C_n$, and $C=\bigcup_{n\geq 1}C_n$ is the target concept class we wish to learn. Note here that even if our initial input is classical, we can efficiently prepare the quantum sample in every iteration. Note here that this step is efficient with or without the assumption of a quantum random access memory. If we assume that we use a QRAM to prepare the uniform superposition over the classical samples, we only incur an additive $O(\sqrt{M}\log M)$ term in the query complexity, thus retaining our quantum speedup. We point out that we are only using the QRAM for state preparation and not as an oracle for search problems, thereby bypassing the issues mentioned in \cite{arunachalam2015robustness}. For a detailed discussion on preparation of quantum samples without a QRAM we refer the reader to \cite{deWolf2020}, in which the authors only assume quantum query access to the training samples.
The definition of weak learning and strong learning generalize straightforwardly to quantum PAC learners and all classical PAC learnable function classes are learnable in the quantum setting.  The sample complexity of quantum and classical PAC-learners too are equal up to constant factors\cite{arunachalam2018optimal}. A quantum PAC learner $A$ has access to several copies of the quantum example. $A$ performs a POVM measurement at the end to obtain a hypothesis $h$ belonging to its associated hypothesis class $\mathcal{H}_A$. As in the earlier works on quantum PAC learning~\cite{arunachalam2017guest,bshouty1998learning}, we also assume the ability to create and query a quantum oracle $O_h$ from an obtained hypothesis $h$.
%We refer the reader to \cite{arunachalam2017guest,bshouty1998learning} for more details on quantum PAC learning. 

%\subsubsection{Boosting with Domain Partitioning Hypothesis}
\subsection{Boosting using Domain Partitioning Hypothesis}
\label{sec:BoostinDomainPartition}
Consider weak learners with hypothesis that partition the domain space $\mathcal{X}$ of the inputs into a set of mutually exclusive and exhaustive blocks $\{\mathcal{X}_1,\mathcal{X}_2,\ldots,\mathcal{X}_C\}$ such that for all $x,x^{\prime}\in \mathcal{X}_j,\;j\in\{1,\ldots,C\}$, we have $h^{\prime}(x)=h^{\prime}(x^{\prime})$. 
Since the prediction is constant for all training samples assigned to a specific partition, we denote the prediction $h^{\prime}$ for the partition $\mathcal{X}_j$ by the constant $\beta_j$. Note that analogous to the definition of $h^{\prime}$, $\mathrm{sign}(\beta_j)$ gives us the prediction for the partition $\mathcal{X}_j$ while $|\beta_j|$ gives us the confidence of the prediction. Now, the task at hand reduces to finding good values of $\beta_j$ for each $\mathcal{X}_j$.
% We outline the RealBoost algorithm in \cref{alg:RealBoost} (with a few minor modifications as proposed by Friedman~\cite{Friedman2000}) in which we calculate the quantity $\beta_j$ by taking the log of the ratio of weighted fraction of examples with different labels. We have included an overview of AdaBoost in the Appendix (see Section~\ref{sec:adaboost}).
% \marginpar{cite ``observed in practice''}
We give an example to foster an intuitive way of thinking about the confidence of predictions in this particular context. Suppose $\beta_j$ is calculated by taking the log of the ratio of weighted fraction of examples with different labels. Consider a partition which contains $100$ samples with label $-1$ and $5$ samples with label $+1$. Then the weighted prediction for that particular partition will be $-1.3$ which means that we predict all samples in this partition to have a $-1$ label with a confidence rating of $1.3$. Another partition which contains $55$ samples with label $+1$ and $45$ samples with label $-1$ will have a weighted prediction of $0.08$. Here we see that because the majority has $+1$ label we assign it to the entire partition, but we do so with a much lower confidence than in the previous case. This shows us that if there is almost an equal number of samples of both labels in a particular domain, then the confidence for predicting either class will be quite low.

\section{Quantum algorithm for boosting}
In this section, we explain the QRealBoost algorithm in detail which is given in \cref{alg:QRealBoost}. 
% While the algorithm tries to perform operations in superposition as much as possible, observe the slight change in the normalization constant in the distribution update rule (see \cref{eq:NewUpdateRule}) as compared to RealBoost in \cref{alg:RealBoost}.
%\subsection{The QRealBoost Algorithm}
The input to our QRealBoost algorithm consists of the weak learner $A$
% parameters $Q$ (sample complexity of $A$), $C$ (the number of domain partitions performed by $A$ and generally assumed to be a constant)
, and a set of $M$ training samples $S$ as copies of the quantum state $\tfrac{1}{\sqrt{M}} \sum_{i \in [M]} \ket{x_i, y_i}$. Since the algorithm is adaptive we can make a worst case guess for $M$ (number of training samples), $Q$ (sample complexity of $A$), $C$ (number of partitions made by $A$), $\gamma$ (bias of $A$), and $T$ (number of iterations of \cref{alg:QRealBoost}); the algorithm will adapt to the intermediate learners which use more optimistic estimates.

If RealBoost(\cref{alg:RealBoost}) computes the distribution $D^t_i$ in the $t$\textsuperscript{th} iteration, QRealBoost estimates it as $\Tilde{D}^t_i$ in the $t$\textsuperscript{th} iteration (additional care is taken since the latter may not actually be a probability distribution). Similarly, RealBoost computes the confidence-rated predictions for the $t$\textsuperscript{th} iteration and $j$\textsuperscript{th} partition $\beta_{j,t}$, which QRealBoost estimates as $\beta^{\prime}_{j,t}$.
% note that if the weak-learner requires $Q$ samples, then we would like to pass $Q$ copies of the above state to it.
% Here, $Q$ is the time complexity of $A$, $C$ refers to the number of domain partitions performed by $A$\footnote{Note that we assume $C=O(1)$ since it is a fairly reasonable assumption that a weak quantum learner can only partition the domain into a constant number of partitions.}. We also assume quantum query access to $M$ training samples $\{(x_1,y_1),\ldots,(x_M,y_M)\}\in\left(\mathcal{X}\times\{-1,+1\}^{M}\right)$.
%There are bounds on $M$ which we discuss later in detail.
% We discuss later the choice of $M$.
%Next we initialize the parameters $\varepsilon$, $\kappa$, and $\Tilde{D}^1$ as given in \cref{alg:QRealBoost}. Finally, 
Following the earlier works on quantum boosting algorithms, we too assume that during an iteration in the outermost for-loop our algorithm, we have quantum query access to the previous hypotheses $h_1,h_2,\ldots,h_{t-1}$ in the form of oracles $O_{h_1}, O_{h_2}, \ldots, O_{h_t}$, respectively, and the confidence-rated predictions $\beta^{\prime}_{j,t}$, for all $j\in\{1,\ldots,C\}$, $t\geq 1$ and $Z^{\prime}_t$ (for $t\geq 1$), which are stored in quantum registers.

\subsection{Explanation of QRealBoost (\cref{alg:QRealBoost})}

\subsubsection{Preparing quantum examples for training {$A$}}
We consider the $t$\textsuperscript{th} iteration of the outermost loop. We first initialize Q copies of $\ket{\phi_0}$ and $2C$ copies of $\ket{\psi_0}$ both set to $\frac{1}{\sqrt{M}}\sum_{i\in [M]}\ket{x_i, y_i, D^{1}_{i}}$. Oracular access to all the previous hypotheses $\{h_1,\ldots, h_{t-1}\}$ can be expressed as $\ket{x_i,y_i}\ket{0}\xrightarrow[]{O_{h_t}}\ket{x_i,y_i}\ket{j^{t}_{i}}$,
% \begin{equation*}
% \end{equation*}
where $j^{t}_{i}=h_t(x_i)$ refers to the domain partition of the $i$\textsuperscript{th} sample at the $t$\textsuperscript{th} iteration. 

We query each such oracle in order to produce $Q+2C$ copies of $\ket{\phi_1}=\ket{\psi_1}=\frac{1}{\sqrt{M}}\sum_{i\in [M]}\ket{x_i, y_i,\Tilde{ D^{1}_{i}}}\ket{j^{1}_{i},j^{2}_{i},\ldots,j^{t-1}_{i}}$. Next, using the stored class weights $\beta^{\prime}_{j,s}$, $Z^{\prime}_{s}$, and the oracles to the hypotheses, we construct a unitary mapping $U_D$ for updating the weight register using $t-1$ applications of \cref{eq:NewUpdateRule} as follows: $\ket{\phi_1}\xrightarrow[]{U_D}\ket{\phi_2}=\frac{1}{\sqrt{M}}\sum_{i\in [M]}\ket{x_i, y_i, \Tilde{D^{t}_{i}}}\ket{j^{1}_{i},j^{2}_{i},\ldots,j^{t-1}_{i}}$. We perform this update too on all $Q$ copies of $\ket{\phi_1}$ and $2C$ copies of $\ket{\psi_1}$.

\subsubsection{Training {$A$} to obtain a new hypothesis}
For all $Q$ copies of $\ket{\phi_2}$, we perform a conditional rotation on the register $\ket{\Tilde{D^{t}_{i}}}$ to obtain the state $\ket{\phi_3}=\sum_{i\in [M]} \frac{1}{\sqrt{M}} \ket{x_i, y_i, \Tilde{D^{t}_{i}}}\ket{j^{1}_{i},\ldots,j^{t-1}_{i}}\left(\sqrt{\Tilde{D^{t}_{i}}}\ket{1}+\sqrt{1-\Tilde{D^{t}_{i}}}\ket{0}\right)$.
% \begin{equation}
% \label{eq:phi3}
%     \ket{\phi_3}=\sum_{i\in [M]} \frac{1}{\sqrt{M}} \ket{x_i, y_i, \Tilde{D^{t}_{i}}}\ket{j^{1}_{i},\ldots,j^{t-1}_{i}}\left(\sqrt{\Tilde{D^{t}_{i}}}\ket{1}+\sqrt{1-\Tilde{D^{t}_{i}}}\ket{0}\right)
% \end{equation}
Let $U_{0\rightarrow3}$ be the unitary that performs $\ket{0}\xrightarrow[]{}\ket{\phi_3}$. We perform Amplitude Amplification as stated in \cref{thm:AmplitudeAmplification} on $\ket{\phi_3}$ to obtain the state $\ket{\phi_4}$ (using $O(\sqrt{M}\log{T})$ applications of $U_{0\rightarrow3}$ and $U_{0\rightarrow3}^{-1}$) with probability at least $O(1-1/T)$. The state $\ket{\phi_4}$ is $\sum_{i\in [M]}\sqrt{\Tilde{D^{t}_{i}}}\ket{x_i, y_i, \Tilde{D^{t}_{i}}}\ket{j^{1}_{i},\ldots,j^{t-1}_{i}}+\ket{\zeta_t}$.
% \begin{equation}
% \label{eq:amplifiedstate}
%     \ket{\phi_4}=\sum_{i\in [M]}\sqrt{\Tilde{D^{t}_{i}}}\ket{x_i, y_i, \Tilde{D^{t}_{i}}}\ket{j^{1}_{i},\ldots,j^{t-1}_{i}}+\ket{\zeta_t}
% \end{equation}

The state $\ket{\zeta_t}$ is present since $\sum_{i\in [M]}{\Tilde{D^{t}_{i}}}\leq 1$ (i.e. the weights are sub-normalized). We state a claim now (see \cref{subsec:proof_bound_on_subnormalized_sum} for proof) that shows that the sum of the weights is very close to 1, and hence, very little interference is expected from $\ket{\zeta_t}$.
\begin{claim}
\label{clm:bound_on_subnormalized_sum}
For $\Tilde{D}^t_i$ updated as given in \cref{eq:NewUpdateRule} and $t\in\{1,\ldots,T\}$, we can bound the sum of the sub-normalized weights as $\sum_{i\in[M]}  \Tilde{D}^{t}_{i}\in\left[1-\frac{4\varepsilon}{1+\varepsilon},1\right]$.
\end{claim}

Now, uncompute by applying $U^{-1}_D$ and $O_{h_1}^{-1}\ldots,O_{h_{t-1}^{-1}}$ to $\ket{\phi_4}$ to obtain the state $\ket{\phi_5}=\sum_{i\in [M]}\sqrt{\Tilde{D^{t}_{i}}}\ket{x_i, y_i}+\ket{\zeta_t^{\prime}}$.
We pass $Q$ copies of $\ket{\phi_5}$ to the weak learner $A$. In turn, the weak learner produces a hypothesis $h_t$ \footnote{\label{foot:note1}We assume that the probability of $A$ not producing any hypothesis is $O(1/T)$, similar to earlier works~\cite{gavinsky2003optimally}.} to which we assume oracular access. The following claim (proof in \cref{subsec:proof_nonidealhypothesis}) shows that the learned hypothesis is a good hypothesis.
\begin{claim}
\label{clm:nonidealhypothesis}
If at the $t$\textsuperscript{th} iteration, the $\gamma$-weak learner $A$ produces a hypothesis $h_t$ on being fed $Q$ copies of the ideal state $\ket{\phi_5^{\prime}}=\sum_{i\in [M]}\sqrt{{D^{t}_{i}}}\ket{x_i, y_i}$, then $A$ produces the same hypothesis $h_t$ with high probability when given $Q$ copies of $\ket{\phi_5}$.
\end{claim}
\subsubsection{Obtaining confidence-rated predictions on sample points}

At this point, we have $2C$ copies of $\ket{\psi_2} = \frac{1}{\sqrt{M}}\sum_{i\in [M]}\ket{x_i, y_i, \Tilde{D^{t}_{i}}}\ket{j^{1}_{i},j^{2}_{i},\ldots,j^{t-1}_{i}}$ (from applying $U_D$ to $\ket{\psi_1}$) and oracular access to the hypothesis $h_t$. We perform the unitary transformation
\begin{equation*}
\label{eq:psi3}
    \begin{split}
        \left[\ket{\psi_2}\ket{0}\right]^{\otimes 2C}\xrightarrow[]{O_{h_t}}\left[\ket{\psi_3}\right]^{\otimes 2C}
        % &=\bigotimes_{\ell=1}^{2C}\left[\frac{1}{\sqrt{M}}\sum_{i\in [M]}\ket{x_i, y_i, \Tilde{D^{t}_{i}}}\ket{j^{1}_{i},j^{2}_{i},\ldots,j^{t-1}_{i}}\ket{{j^{t}_{i}}}\right]\\
        &=\bigotimes_{\substack{k\in\{1,\ldots,C\}\\b\in\{-1,+1\}}}\left[\frac{1}{\sqrt{M}}\sum_{i\in [M]}\ket{x_i, y_i, \Tilde{D^{t}_{i}}}\ket{j^{1}_{i},j^{2}_{i},\ldots,j^{t-1}_{i}}\ket{{j^{t}_{i}}}\right]
    \end{split}
\end{equation*}
Consider the $(k,b)$\textsuperscript{th} copy of $\left[\ket{\psi_3}\right]^{\otimes 2C}$ for $k\in\{1,2,\ldots,C\}$ and $b\in\{-1,+1\}$. Perform the update  $\psi_3\xrightarrow[]{}\psi_4$ as
\begin{equation*}
\label{eq:psi_4}
    \ket{\psi_3}_{(k,b)}\left(\ket{0}^{\otimes 2}\right)\xrightarrow[]{}\ket{\psi_4}_{(k,b)}=\frac{1}{\sqrt{M}}\sum_{i\in[M]}\ket{x_i}\ket{y_i}\ket{\Tilde{D}_i^{t}}\underbrace{\ket{j^{1}_{i},\ldots,j^{t}_{i}}}_{\ket{j(i,t)}}\underbrace{\ket{\mathbb{I}[j_i^{t}=k]}}_{{{\ket{\mathbb{I}_1}}}}\underbrace{\ket{\mathbb{I}[y_i=b]}}_{{{\ket{\mathbb{I}_2}}}}.
\end{equation*}
Note here that $\mathbb{I}_1$ and $\mathbb{I}_2$ are binary valued states. 
% Also note that we do not need any complicated computations for this step. Let the binary representation of $j^{t}_{i}$ be $j^{t}_{i,1}j^{t}_{i,2}\ldots j^{t}_{i,n}$. Similarly, let $k=k_1k_2\ldots k_n$. We carry out the following unitary transformation
% \begin{equation*}
%     \ket{j^{t}_{i,1},\ldots, j^{t}_{i,n}}\ket{k_1,\ldots,k_n}\ket{0^{\otimes{n}}}\rightarrow\ket{j^{t}_{i,1},\ldots, j^{t}_{i,n}}\ket{k_1,\ldots,k_n}\ket{j^{t}_{i,1}\cdot k_1,\ldots,j^{t}_{i,n}\cdot k_n}
% \end{equation*}
% and use the individual bits of the register $\ket{j^{t}_{i,1}\cdot k_1,\ldots,j^{t}_{i,n}\cdot k_n}$ to set an ancilla bit $\ket{\mathbb{I}_1}$ which serves as the indicator register for the function $[j^t_i=k]$. $\ket{\mathbb{I}_2}$ is obtained similarly. 
Using $\ket{\mathbb{I}_1}$ and $\ket{\mathbb{I}_2}$ as controls, we obtain the state
\begin{equation*}
\label{eq:psi_5}
    \ket{\psi_5}_{(k,b)}=\frac{1}{\sqrt{M}}\sum_{i\in[M]}\ket{x_i}\ket{y_i}\ket{\Tilde{D}_i^{t}}\ket{j(i,t)}\ket{\mathbb{I}_1}\ket{\mathbb{I}_2}\underbrace{{\ket{\Tilde{D}_i^{t}\cdot \mathbb{I}_1 \cdot \mathbb{I}_2}}}_{\Tilde{D}^{k,b,t}_{i}}
\end{equation*}
Now we perform a conditional rotation on the $\ket{\Tilde{D}^{k,b,t}_{i}}$ register to obtain~\footnote{ Here 
\mbox{${W}^{k,t}_{b}={\sum_{i\in[M]}{\Tilde{D}^{k,b,t}_{i}}}/{M}$}, \mbox{$\ket{\chi}^{1}_{(k,b)} = \frac{1}{\sqrt{M}}\sum_{i\in[M]}\frac{\sqrt{\Tilde{D}^{k,b,t}_{i}}}{\sqrt{{W}^{k,t}_{b}}}\ket{x_i}\ket{y_i}\ket{\Tilde{D}_i^{t}}\ket{j(i,t)}\ket{\mathbb{I}_1}\ket{\mathbb{I}_2}{\ket{\Tilde{D}^{k,b,t}_{i}}}$} and \mbox{$
    \ket{\chi}^{0}_{(k,b)} = \frac{1}{\sqrt{M}}\sum_{i\in[M]}\frac{\sqrt{1-\Tilde{D}^{k,b,t}_{i}}}{\sqrt{1-{W}^{k,t}_{b}}}\ket{x_i}\ket{y_i}\ket{\Tilde{D}_i^{t}}\ket{j(i,t)}\ket{\mathbb{I}_1}\ket{\mathbb{I}_2}{\ket{\Tilde{D}^{k,b,t}_{i}}}
$} }
\begin{equation*}
\label{eq:psi_6}
    \begin{split}
        \ket{\psi_6}_{(k,b)}
        &=\sqrt{{W}^{k,t}_{b}}\ket{\chi}^{1}_{(k,b)}\ket{1}+\sqrt{1-{{W}^{k,t}_{b}}}\ket{\chi}^{0}_{(k,b)}\ket{0}
    \end{split}
\end{equation*}

Let $V_{0,6}^{(k,b)}$ be the unitary that performs $\ket{0}\xrightarrow[]{}\ket{\psi_6}_{(k,b)}$. We perform relative-error amplitude estimation as stated in \cref{thm:AmplitudeEstimation}, with an expected $\Tilde{O}(\sqrt{M}QT^{2})$ queries to $V_{0,6}^{(k,b)}$ and $V_{0,6}^{-1\;\;(k,b)}$ to obtain the quantity $\Tilde{W}^{k,t}_{b}$ that is an estimate of $W^{k,t}_b$ with high probability. 
% To obtain the above result, we set $p=O\left(\frac{1}{M}\right)$ and $\varepsilon=\frac{1}{QT^2}$. 
We carry out relative error amplitude estimation as stated in \cref{thm:AmplitudeEstimation} to estimate the quantity $W^{k,t}_{b}$. 

Hence, we obtain all $2C$ values of $\Tilde{W}^{j,t}_{b}$ for all $j\in\{1,2,\ldots,C\}$, $b\in\{-1,+1\}$. Note that it is possible for the value of $W^{j,t}_{b}$ to be very small (even zero) for some $j$. This would result in the quantities $\beta^{\prime}_{j,t}$ becoming very large or unbounded, thus increasing the tendency of the learner to overfit. We use a general smoothing technique known as \textit{Laplace correction}\cite{clark1989cn2} to overcome this issue\footnote{The details are explained in \cref{sec:laplace}.}
% Note that this step can also be incorporated into the classical RealBoost algorithm since if the weights were normalized before smoothing, they are still normalized post smoothing. 
%Hence, we obtain all $2C$ values of $\Tilde{W}^{j,t}_{b}$ for all $j\in\{1,2,\ldots,C\}$ and perform smoothing. 
and use the smoothed values to calculate the margins as $\beta^{\prime}_{j,t}=\frac{1}{2}\ln{\left(\frac{\Tilde{W}^{j,t}_{+}}{\Tilde{W}^{j,t}_{-}}\right)}\;\;\;\;\;\;\;\forall j\in\{1,\ldots,C\}$
and the normalization constant as $Z^{\prime}_t=2\sum_{j=1}^{C}\sqrt{\Tilde{W}^{j,t}_{+}\Tilde{W}^{j,t}_{-}}$.
% \begin{equation}
% \label{eq:zed_est}
%     Z^{\prime}_t=2\sum_{j=1}^{C}\sqrt{\Tilde{W}^{j,t}_{+}\Tilde{W}^{j,t}_{-}}.
% \end{equation}
At this point, we can make the following two claims.
\begin{claim}
\label{clm:marginbound}
Let the weights be relatively estimated using the error parameter $\varepsilon$ by \cref{alg:QRealBoost}; i.e. $\left|W^{j,t}_{b}-\Tilde{W}^{j,t}_{b}\right|\leq \varepsilon\cdot W^{j,t}_{b}$. Then the difference between the actual margins $\beta_{j,t}$  and the estimated margins $\beta^{\prime}_{j,t}$ is bounded as $\left|\beta^{\prime}_{j,t}-\beta_{j,t}\right|\leq \frac{1}{2}\ln{\left(\frac{1+\varepsilon}{1-\varepsilon}\right)}; \;\;\;\; j\in\{1,2,\ldots,C\}$.
% \begin{equation}
% \label{eq:marginbound}
%     \left|\beta^{\prime}_{j,t}-\beta_{j,t}\right|\leq \frac{1}{2}\ln{\left(\frac{1+\varepsilon}{1-\varepsilon}\right)}; \;\;\;\; j\in\{1,2,\ldots,C\}
% \end{equation}
\end{claim}
\begin{claim}
\label{clm:normconstbound}
In the same setting as in \cref{clm:marginbound}, 
%Let the weights be relatively estimated using the error parameter $\varepsilon$ by \cref{alg:QRealBoost}; i.e. $\left|W^{j,t}_{b}-\Tilde{W}^{j,t}_{b}\right|\leq \varepsilon\cdot W^{j,t}_{b}$. Then, 
the deviation in the normalization constant at every iteration is bounded as $\left|Z^{\prime}_{t}- Z_t\right|\leq \varepsilon\cdot Z_t$.
\end{claim}
We present the proof of \cref{clm:marginbound} and \cref{clm:normconstbound} in \cref{subsec:proof_marginbound} and \cref{subsec:proof_normconstbound} respectively. From \cref{clm:marginbound} we see that the difference between the actual margin and the estimated margins is very small. In fact, a very simple calculation shows us that $\left|\beta^{\prime}_{j,t}-\beta_{j,t}\right|\leq 0.1$ for $\varepsilon\leq 0.1$. We note that the error parameter $\varepsilon$ is far smaller than a constant fraction, which means our estimated margins are quite close to the ideal margin values. \cref{clm:normconstbound} shows that when we minimize the normalization constant at every step using the estimated values $\Tilde{W}^{j,t}_{b}$, these quantities are themselves relatively bounded by the actual normalization constant. This implies that the training error of the combined classifier is greedily minimized when the normalization constant is minimized at every step\footnote{See \cref{thm:trainerr_bound}.}. Hence, our training error at every step does not blow up due to estimation of the partition weights. Now, we plug in the values of $\kappa$ (as initialized in \cref{alg:QRealBoost}), $\beta^{\prime}_{j,t}$, and $Z^{\prime}_{t}$ in to \cref{eq:NewUpdateRule} to perform the update from $\Tilde{D}^t_{i}$ to $\Tilde{D}^{t+1}_{i}$ for all $i\in[M]$. The output of the algorithm is the final hypothesis $H(x)=\mathrm{sign}\left(\sum_{t=1}^{T}\beta^{\prime}_{j,t}\right)$ where our weak learner assigns any training example $x\sim D$ the $j$\textsuperscript{th} partition at the $t$\textsuperscript{th} iteration, and $\beta^{\prime}_{j,t}$ is the weighted prediction of the $j$\textsuperscript{th} partition at the $t$\textsuperscript{th} iteration.
\subsection{Proof of correctness}
The probability of failure of \cref{alg:QRealBoost} stems primarily from the steps \ref{step1}, \ref{step2}, and \ref{step3}, where each step fails with a probability at most $O(1/T)$. When we take a union bound over all $T$ iterations for all three steps, the overall failure probability dips to an arbitrary constant which is at most $1/3$. There is an extra log factor incurred due to error reduction which can be absorbed in the $\Tilde{O}(.)$ notation. We now state the following claim regarding the training error of our algorithm the proof of which is included in \cref{subsec:proof_zerotrainingerror}.
\begin{claim}
\label{clm:zerotrainingerror}
For a sufficiently large number of iterations $T\geq\frac{\ln{M}}{2\gamma^2}$, our combined classifier $H$ has zero training error with respect to the uniform superposition $\Tilde{D}^1$ with high probability.
\end{claim}
From \cref{clm:zerotrainingerror} and \cref{corr:generalizationsamples}, we also conclude that if we run \cref{alg:QRealBoost} for a sufficiently large number of iterations $T$, then with a high probability we output a hypothesis $H$ according to \cref{eq:qrealboostcombinedclassifier} that has zero training error and a small generalization error.
\subsection{Complexity Analysis}
In this section we state the query complexity and the time complexity of our algorithm.
\begin{theorem}[Query Complexity]\label{thm:quercomp}
Suppose we boost a $\gamma$-weak learner $A$ with sample complexity $Q$, and an associated hypothesis class $\mathcal{H}$ having VC dimension $d_{\mathcal{H}}$ using \cref{alg:QRealBoost}. If the weak learner $A$ produces at most $C$ partitions at every iteration, then the query complexity of \cref{alg:QRealBoost} is $O\left(\frac{\sqrt{d_{\mathcal{H}}}\cdot C\cdot Q}{\gamma^{9}}\right)$.
\end{theorem}
\begin{theorem}[Time Complexity]\label{thm:timecomp}
Suppose we boost a $\gamma$-weak learner $A$ with sample complexity $Q$, and an associated hypothesis class $\mathcal{H}$ having VC dimension $d_{\mathcal{H}}$ using \cref{alg:QRealBoost}. The size of the class $\mathcal{C}$ is assumed to be $n$. If the weak learner $A$ produces at most $C$ partitions at every iteration, then the time complexity of \cref{alg:QRealBoost} is $O\left(\frac{n^2\sqrt{d_{\mathcal{H}}} C Q}{\gamma^{9}}\right)$
\end{theorem}
The proof of \cref{thm:quercomp} is given in \cref{sec:querycomp} and the proof of \cref{thm:timecomp} is given in \cref{sec:timecomp}.
\vspace*{-2pt}

\section{Experiments}

\begin{figure}
    \centering
    \includegraphics[width = 0.7\textwidth]{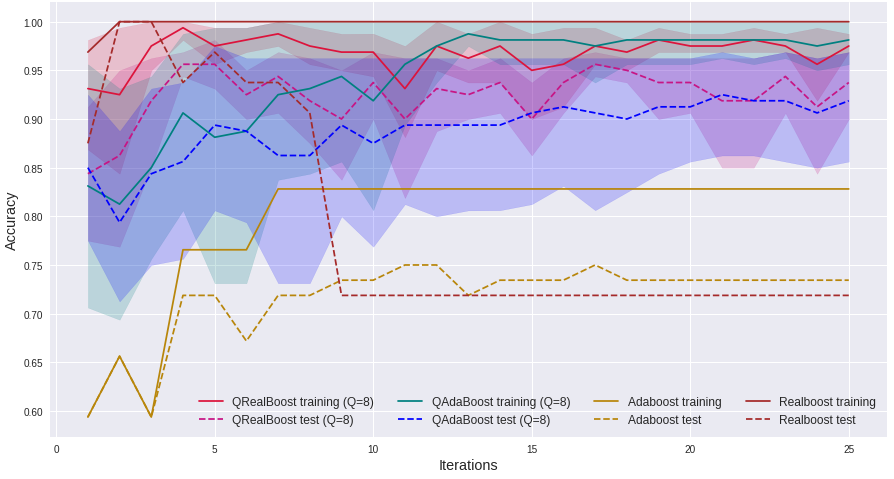}
    \caption{{\footnotesize Comparing the performance of 4 different boosting algorithms using $k$-means (with $k=3$) as the weak learner on the \href{http://yann.lecun.com/exdb/mnist/}{MNIST dataset} for the digits $4$ and $5$ using 32 training samples. This experiment retains the binary classification nature of the problem.}}
    \label{fig:mnist}
\end{figure}

We compare\footnote{Our code is freely available at \url{https://github.com/braqiiit/QRealBoost}.} the generalization ability and convergence of QRealBoost against AdaBoost, RealBoost, and QAdaBoost on the Breast Cancer Wisconsin dataset( see~\cref{fig:breastcancer}) and the MNIST dataset (see~\cref{fig:mnist}). Since there does not exist any quantum simulators or actual quantum backends large enough to test QRealBoost, we had to make some interesting choices and changes in the implementation which are detailed below. 
\begin{enumerate}
    \item We focus on qualitative analysis behaviour (training and test convergence) of the algorithms in these experiments rather than its efficiency due to the lack of quantum simulators and quantum backends with a sufficient number of qubits.
    \item Instead of computing the distribution weights from scratch, we store the updated distribution weights after every iteration. This is done because in the former approach the number of qubits needed to store the weights up to a reasonable degree of precision blows up with the number of iterations, taking even experiments with few training samples out of our reach. Even though this choice sacrifices the quantum speedup it does not affect the convergence behaviour of QRealBoost.
    \item We used a classical weak learner ($k$-means) since off-the-shelf quantum weak learners are not readily available right now, and implementing one was out of the scope of this work. The implementation can be easily modified to use any kind of learner implemented as a quantum circuit. We measure the $\ket{\phi_5}$ state and pass the top $Q$ training samples to the $k$-means algorithm. In \cref{sec:quantumpaclearning} we pointed out that this is exactly how quantum learners could simulate classical learners.
    \item We use the the \textit{IterativeAmplitudeEstimation} class provided by Qiskit which is an implementation of the Iterative Quantum Amplitude Estimation (IQAE) algorithm~\cite{grinko2021iterative} that replaces quantum phase estimation with clever use of Grover iterations. Our choice was motivated by the availability and performance of the algorithm which helped us decrease the number of qubits needed for the implementation. An important note is the fact that even though the experiments were conducted with additive estimation instead of relative estimation, we still managed to boost the weak learner.
\end{enumerate}

The lines for QAdaBoost and QRealBoost represent a mean accuracy over $5$ independent experiments, while the hue bands represent the standard deviation across all experiments. The QAdaBoost and QRealBoost algorithms are tested on quantum simulators (instead of actual quantum backends) due to quantum resource limitations. We set the sample complexity of QAdaBoost and QRealBoost to be $8$ for both sets of experiments. All algorithms are trained on $32$ samples for both sets of experiments, and we observe the training accuracy and test accuracy over $25$ iterations. 

In the first experiment (see~\cref{fig:breastcancer}), we observe that both QRealBoost and QAdaBoost have similar convergence rates {\it w.r.t.} training error that is better than RealBoost and completely dwarfs AdaBoost. Moreover, QRealBoost converges faster than QAdaBoost and has a tighter deviation in training loss over five experiments, especially in the early iterations. Even in terms of generalization ability, QRealBoost completely outperforms QAdaBoost and RealBoost and is only surpassed by AdaBoost. In the second experiment (see~\cref{fig:mnist}), RealBoost appears to overfit the training samples and suffers from the worst generalization error out of all four algorithms. AdaBoost has a poor convergence rate and generalization error as well. QRealBoost and QAdaBoost perform similarly in training accuracy, with QRealBoost narrowly beating QAdaBoost via faster convergence and a tighter deviation. Regarding generalization abilities, QAdaBoost loses out to QRealBoost in overall test accuracy and deviation over experiments, albeit with a much smaller margin. These are encouraging observations, especially considering that QRealBoost trains on $8$ samples at every iteration, while the classical algorithms have access to all the $32$ samples every iteration.

\vspace*{-2pt}
\section{Conclusions}
In this work, we designed the QRealBoost algorithm which tackles an open question posed by Izdebski et al.~\cite{deWolf2020} to boost weak quantum PAC learners that output non-binary hypotheses. QRealBoost retains the performance of RealBoost which has superior theoretical properties (supported by empirical evidence too) as compared to the celebrated boosting algorithm, AdaBoost~\cite{Freund1997}. We also establish that both theoretically and empirically QRealBoost outperforms QAdaBoost which is the only other known adaptive quantum boosting algorithm.

An issue with QRealBoost is complexity of $\gamma$, arising from recomputing $\Tilde{D}^t$ over the training samples at every iteration from scratch. We believe that this computation can be avoided by maintaining a ``distribution oracle'' which only needs to be updated in each iteration. If it turns out that the lower bound on $\gamma$ is worse for quantum boosting algorithms compared to classical boosting algorithms in the general case, the next question would be finding (or even determining the existence of) relevant hypotheses classes in which quantum boosting provides us with an advantage.

We also observe that the constant factor $C$ in the numerator of the time complexity may be exponentially reduced by simultaneously estimating the individual domain partition weights using amplitude estimation techniques as shown in \cite{van2021quantum}, and this is a possible direction of future work. 

A logical continuation of this work is quantizing other variants of AdaBoost which depend on domain partitioning hypotheses such as GentleBoost\cite{Friedman2000}, ModestBoost\cite{vezhnevets2005modest}, Parameterized AdaBoost\cite{wu2014parameterized}, and Penalized AdaBoost\cite{wu2015penalized}. Each variant has different generalization abilities, which make them useful in different contexts. The algorithmic framework followed in this work for estimating the partition weights may be useful to model quantum versions of these variants.

\bibliography{refs.bib}

\appendix

\section{Proofs}
In this section, we present the proofs of the claims made in the earlier section. First, we restate some well-known quantum subroutines that we use throughout this work to prove our main results.

\subsection{Quantum subroutines for amplitude amplification and estimation}

\begin{theorem}[Amplitude Amplification \cite{MR1947332}]
\label{thm:AmplitudeAmplification}
 Let there be a unitary $U$ such that $U\ket{0}=\sqrt{a}\ket{\phi_0}+\sqrt{1-a}\ket{\phi_1}$ for an unknown $a>0$. If $a>p>0$, then there exists a quantum amplitude amplification algorithm that outputs the state $\ket{\phi_0}$ with a probability $p^{\prime}>0$. The expected number of calls to $U$ and $U^{-1}$ made by our quantum amplitude amplification algorithm is $\Theta(\sqrt{p^{\prime}/p})$.
\end{theorem}
\begin{theorem}[Relative Error Estimation \cite{ambainis:hal-00214301}]
\label{thm:AmplitudeEstimation}
Given an error parameter $\varepsilon$, a constant $k\geq1$, and a unitary $U$ such that $U\ket{0}$ outputs $1$ with probability at least $p$. Then there exists a quantum amplitude estimation algorithm that produces an estimate $\Tilde{a}$ of the success probability $a$ with probability at least $1-\frac{1}{2^k}$ such that $\left|a-\Tilde{a}\right|\leq \varepsilon a$ where $a\geq p$. The expected number of calls to $U$ and $U^{-1}$ made by our quantum amplitude estimation algorithm is
\begin{equation}
    O\left(\frac{k}{\varepsilon\sqrt{p}}\left(1+\log{\log{\frac{1}{p}}}\right)\right)
\end{equation}
\end{theorem}

\subsection[Proof of bound on sub-normalized sum]{Proof of \cref{clm:bound_on_subnormalized_sum} (bound on subnormalized sum)}\label{subsec:proof_bound_on_subnormalized_sum}
We restate the new distribution update rule as given in \cref{alg:QRealBoost}
\begin{equation}
    \Tilde{D}^{t+1}_{i} = \frac{\Tilde{D}^{t}_{i}\mathrm{exp}\left(-\beta^{\prime}_{j,t}\cdot y_i\right)}{\kappa Z^{\prime}_{t}}
\end{equation}
Before getting into the actual proof, we make two observations. We can see from $\ket{\phi_5}$ that our weights are sub-normalised. This gives us a trivial upper bound
\begin{equation}
\label{eq:subnormal_trivialbound}
    \sum_{i\in[M]} \Tilde{D}^{t}_{i} \leq 1;\;\;\;\;\forall t\geq 1
\end{equation}
We now make our second observation:
\begin{equation}
\label{eq:normalizedupdate1}
    \frac{\sum_{i\in[M]}\Tilde{D}^{t}_{i}\mathrm{exp}\left(-\beta^{\prime}_{j,t}\cdot y_i\right)}{\sum_{j=1}^{C}\left(W^{j,t}_{+}\mathrm{exp}\left(-\beta^{\prime}_{j,t}\right)+W^{j,t}_{-}\mathrm{exp}\left(\beta^{\prime}_{j,t}\right)\right)}=1.
\end{equation}
We can arrive at the observation in \eqref{eq:normalizedupdate1} by following the arguments in \textbf{Schapire-Singer}. Let us start by obtaining a preliminary bound on the quantity $\sum_{i} \Tilde{D}^{t+1}_{i}$ as
\begin{equation}
    \begin{split}
        &\phantom{=}\frac{\sum_{i\in[M]}\Tilde{D}^{t}_{i}\mathrm{exp}\left(-\beta^{\prime}_{j,t}\cdot y_i\right)}{\kappa Z^{\prime}_{t}}\\
        &=\frac{\sum_{i\in[M]}\Tilde{D}^{t}_{i}\mathrm{exp}\left(-\beta^{\prime}_{j,t}\cdot y_i\right)}{\sum_{j=1}^{C}\left(W^{j,t}_{+}\mathrm{exp}\left(-\beta^{\prime}_{j,t}\right)+W^{j,t}_{-}\mathrm{exp}\left(\beta^{\prime}_{j,t}\right)\right)}\cdot \frac{\sum_{j=1}^{C}\left(W^{j,t}_{+}\mathrm{exp}\left(-\beta^{\prime}_{j,t}\right)+W^{j,t}_{-}\mathrm{exp}\left(\beta^{\prime}_{j,t}\right)\right)}{\kappa Z^{\prime}_{t}}\\
        &=\frac{\sum_{j=1}^{C}\left(W^{j,t}_{+}\mathrm{exp}\left(-\beta^{\prime}_{j,t}\right)+W^{j,t}_{-}\mathrm{exp}\left(\beta^{\prime}_{j,t}\right)\right)}{\kappa Z^{\prime}_{t}}\\
        &=\frac{\sum_{j=1}^{C}\left(W^{j,t}_{+}\mathrm{exp}\left(-\beta^{\prime}_{j,t}\right)+W^{j,t}_{-}\mathrm{exp}\left(\beta^{\prime}_{j,t}\right)\right)}{ 2\kappa\sum_{j=1}^{C}\sqrt{\Tilde{W}^{j,t}_{+}\cdot \Tilde{W}^{j,t}_{-}}}\\
        &=\frac{1}{2\kappa}\sum_{j=1}^{C}\frac{W^{j,t}_{+}\mathrm{exp}\left(-\beta^{\prime}_{j,t}\right)+W^{j,t}_{-}\mathrm{exp}\left(\beta^{\prime}_{j,t}\right)}{\sqrt{\Tilde{W}^{j,t}_{+}\cdot \Tilde{W}^{j,t}_{-}}}
    \end{split}
\end{equation}
The first equality follows from plugging in \cref{eq:NewUpdateRule}. The third equality follows from \cref{eq:normalizedupdate1}. In the fourth, equality we use the value of $Z^{\prime}_{t}$ given in \cref{alg:QRealBoost}. Now we upperbound and lowerbound the quantity $\sum_{i} \Tilde{D}^{t+1}_{i}$ by plugging in \eqref{eq:relative_error_wjb} as
\begin{equation}
\label{eq:sumupperbound}
        \sum_{i\in[M]} \Tilde{D}^{t+1}_{i}\leq\frac{1}{2\kappa\left(1-\varepsilon\right)}\sum_{j=1}^{C}\frac{W^{j,t}_{+}\mathrm{exp}\left(-\beta^{\prime}_{j,t}\right)+W^{j,t}_{-}\mathrm{exp}\left(\beta^{\prime}_{j,t}\right)}{\sqrt{{W}^{j,t}_{+}\cdot {W}^{j,t}_{-}}}
\end{equation}
\begin{equation}
\label{eq:sumlowerbound}
        \sum_{i\in[M]} \Tilde{D}^{t+1}_{i}\geq\frac{1}{2\kappa\left(1+\varepsilon\right)}\sum_{j=1}^{C}\frac{W^{j,t}_{+}\mathrm{exp}\left(-\beta^{\prime}_{j,t}\right)+W^{j,t}_{-}\mathrm{exp}\left(\beta^{\prime}_{j,t}\right)}{\sqrt{{W}^{j,t}_{+}\cdot {W}^{j,t}_{-}}}
\end{equation}
Substituting $\kappa=\frac{C}{\left(1-\varepsilon\right)}\sqrt{\frac{1+\varepsilon}{1-\varepsilon}}$ in \cref{eq:sumupperbound} we have
\begin{equation}
\label{eq:subnormalupbnd}
    \begin{split}
        % &\frac{1}{2\kappa\left(1-\varepsilon\right)}\sum_{j=1}^{C}\frac{W^{j,t}_{+}\mathrm{exp}\left(-\beta^{\prime}_{j,t}\right)+W^{j,t}_{-}\mathrm{exp}\left(\beta^{\prime}_{j,t}\right)}{\sqrt{{W}^{j,t}_{+}\cdot {W}^{j,t}_{-}}}\leq 1\\
        % \implies &\frac{1}{2\kappa\left(1-\varepsilon\right)}\sum_{j=1}^{C}\sqrt{\frac{{W^{j,t}_{+}}}{{W^{j,t}_{-}}}}\cdot\mathrm{exp}\left(-\beta^{\prime}_{j,t}\right)+\sqrt{\frac{{W^{j,t}_{-}}}{{W^{j,t}_{+}}}}\cdot\mathrm{exp}\left(\beta^{\prime}_{j,t}\right)\leq 1\\
        % \implies &\frac{1}{2\kappa\left(1-\varepsilon\right)}\sum_{j=1}^{C}\sqrt{\frac{{W^{j,t}_{+}}}{{W^{j,t}_{-}}}}\cdot\sqrt{\frac{{\Tilde{W}^{j,t}_{-}}}{{\Tilde{W}^{j,t}_{+}}}}+\sqrt{\frac{{W^{j,t}_{-}}}{{W^{j,t}_{+}}}}\cdot\sqrt{\frac{{\Tilde{W}^{j,t}_{+}}}{{\Tilde{W}^{j,t}_{-}}}}\leq 1\\
        % \implies &\frac{1}{2\kappa\left(1-\varepsilon\right)}\sum_{j=1}^{C}\left(2\sqrt{\frac{1+\varepsilon}{1-\varepsilon}}\right)\leq 1\\
        % \implies&\frac{C}{\kappa\left(1-\varepsilon\right)}\sqrt{\frac{1+\varepsilon}{1-\varepsilon}}\leq 1\\
        % \implies&\kappa\geq\frac{C}{\left(1-\varepsilon\right)}\sqrt{\frac{1+\varepsilon}{1-\varepsilon}}
         \sum_{i\in[M]} \Tilde{D}^{t+1}_{i}&\leq\frac{1}{2\kappa\left(1-\varepsilon\right)}\sum_{j=1}^{C}\frac{W^{j,t}_{+}\mathrm{exp}\left(-\beta^{\prime}_{j,t}\right)+W^{j,t}_{-}\mathrm{exp}\left(\beta^{\prime}_{j,t}\right)}{\sqrt{{W}^{j,t}_{+}\cdot {W}^{j,t}_{-}}}\\
        &=\frac{1}{2\kappa\left(1-\varepsilon\right)}\sum_{j=1}^{C}\sqrt{\frac{{W^{j,t}_{+}}}{{W^{j,t}_{-}}}}\cdot\sqrt{\frac{{\Tilde{W}^{j,t}_{-}}}{{\Tilde{W}^{j,t}_{+}}}}+\sqrt{\frac{{W^{j,t}_{-}}}{{W^{j,t}_{+}}}}\cdot\sqrt{\frac{{\Tilde{W}^{j,t}_{+}}}{{\Tilde{W}^{j,t}_{-}}}}\\
        &\leq \frac{1}{2\kappa\left(1-\varepsilon\right)}\sum_{j=1}^{C}\left(2\sqrt{\frac{1+\varepsilon}{1-\varepsilon}}\right)\\
        &= \frac{C}{\kappa\left(1-\varepsilon\right)}\sqrt{\frac{1+\varepsilon}{1-\varepsilon}}=1
    \end{split}
\end{equation}
Similarly, substituting $\kappa=\frac{C}{\left(1-\varepsilon\right)}\sqrt{\frac{1+\varepsilon}{1-\varepsilon}}$ in \cref{eq:sumlowerbound} we have
\begin{equation}
\label{eq:subnormallbnd}
    \begin{split}
        \sum_{i\in[M]} \Tilde{D}^{t+1}_{i}&\geq\frac{1}{2\kappa\left(1+\varepsilon\right)}\sum_{j=1}^{C}\frac{W^{j,t}_{+}\mathrm{exp}\left(-\beta^{\prime}_{j,t}\right)+W^{j,t}_{-}\mathrm{exp}\left(\beta^{\prime}_{j,t}\right)}{\sqrt{{W}^{j,t}_{+}\cdot {W}^{j,t}_{-}}}\\
        &=\frac{1}{2\kappa\left(1+\varepsilon\right)}\sum_{j=1}^{C}\sqrt{\frac{{W^{j,t}_{+}}}{{W^{j,t}_{-}}}}\cdot\sqrt{\frac{{\Tilde{W}^{j,t}_{-}}}{{\Tilde{W}^{j,t}_{+}}}}+\sqrt{\frac{{W^{j,t}_{-}}}{{W^{j,t}_{+}}}}\cdot\sqrt{\frac{{\Tilde{W}^{j,t}_{+}}}{{\Tilde{W}^{j,t}_{-}}}}\\
        &\geq \frac{C}{\kappa\left(1+\varepsilon\right)}\sqrt{\frac{1-\varepsilon}{1+\varepsilon}}=\left(\frac{1-\varepsilon}{1+\varepsilon} \right)^{2}=\left(1-\frac{2\varepsilon}{1+\varepsilon} \right)^{2}\\&\geq 1-\frac{4\varepsilon}{1+\varepsilon}
        % >1-4\varepsilon
    \end{split}
\end{equation}
Combining \cref{eq:subnormalupbnd} and \cref{eq:subnormallbnd} we have for any $t = 1,2, \ldots, T$
\begin{equation}
     \sum_{i\in[M]} \Tilde{D}^{t}_{i}\in\left[1-\frac{4\varepsilon}{1+\varepsilon},1\right]
\end{equation}

\subsection[Proof of hypothesis learning]{Proof of \cref{clm:nonidealhypothesis} (hypotheses are learned correctly {\it w.h.p.})}\label{subsec:proof_nonidealhypothesis}
Let us assume that when we supply $Q$ copies of the state $\sum_{i\in[M]} \sqrt{D^{t}_{i}}\ket{x_i,y_i}$ we obtain the hypothesis $h_t$ with probability $\rho$. We want to bound the probability $\sigma$ of obtaining the same hypothesis $h_t$ when we give our weak learner $A$ the state $\phi_5$. Before we dive into the calculations, we define a few terms.
\begin{definition}[Fidelity]
Fidelity is a measure of the closeness of two quantum states. When we have two pure states $\ket{\psi}$ and $\ket{\phi}$ we define Fidelity between the two states as
\begin{equation}
    F({\psi},{\phi})=F({\phi},{\psi})=\left|\bra{\psi}\ket{\phi}\right|^2
\end{equation}
% $$
% F({\psi},{\phi})=F({\phi},{\psi})=\left|\bra{\psi}\ket{\phi}\right|^2
% $$
Let $\rho$ and $\sigma$ be the density matrices of $\psi$ and $\phi$ respectively. An alternate characterization of Fidelity in terms of density matrices is
\begin{equation}
    F(\rho,\sigma)=\left\Vert\sqrt{\rho}\sqrt{\sigma}\right\Vert_{1}
\end{equation}
\end{definition}
\begin{definition}[Normalized Trace Distance]
Trace distance is another measure of closeness between two quantum states. If there is a set of POVMs $\{E\}$, then the POVM leading to the largest difference in measurement outcomes between two quantum states is the trace distance.
\begin{equation}
    D(\rho,\sigma)=\frac{1}{2}\left\Vert\rho-\sigma\right\Vert_{1}=\underset{E_i}{\mathrm{max}}\;\sum_{i}\left|\mathrm{Tr}\{E(\rho-\sigma)\}\right|
\end{equation}
When $\rho$ and $\sigma$ are density matrices of pure states, Trace distance is related to Fidelity as follows:
\begin{equation}
    D(\rho,\sigma)=\sqrt{1-\mathbb{F}(\rho,\sigma)}
\end{equation}
\end{definition}
Let $p$ be the probability that $A$ outputs the hypothesis $h_t$ on being fed $Q$ copies of the ideal state $\ket{\phi_5^{\prime}}=\sum_{i\in [M]}\sqrt{{D^{t}_{i}}}\ket{x_i, y_i}$. Let $q$ be the probability that $A$ outputs the hypothesis $h_t$ on being fed $Q$ copies of the state $\ket{\phi_5}$. We want to bound the quantity $|p-q|$ and show that this is a small quantity. We denote the class of POVMs on the hypothesis space $\mathcal{H}$ as $\{E_h\}_{h\in \mathcal{H}}$ such that $\sum_{h\in \mathcal{H}}E_h=I$. Then by the above definitions of trace distance and fidelity we have
\begin{equation}\label{eq:fidelity-eqn}
    \begin{split}
        |p-q|&\leq \underset{\{E_h\}}{\mathrm{max}}\left|\mathrm{Tr}\{E_h(\rho-\sigma)\}\right|\leq\sum_{h\in\mathcal{H}}\left|\mathrm{Tr}\{E_h(\rho-\sigma)\}\right|\\
        &=D(\rho-\sigma)=\sqrt{1-\mathbb{F}(\rho,\sigma)}=\left({1-\left|\left(\bra{\phi_5}\ket{\phi_5^{\prime}}\right)^Q\right|^2}\right)^{\frac{1}{2}}\\
        &\leq\left({1-\left|\bra{\phi_5}\ket{\phi_5^{\prime}}\right|^2Q}\right)^{\frac{1}{2}}
    \end{split}
\end{equation}
Now we bound the quantity $\left|\bra{\phi_5}\ket{\phi_5^{\prime}}\right|$.
\begin{equation}
    \left|\bra{\phi_5}\ket{\phi_5^{\prime}}\right|=\left|\sqrt{\Tilde{D}^{t}_{i}\cdot {D}^{t}_{i} } + \bra{\zeta_t}\ket{\phi^{\prime}_{5}}\right|
        \geq\left|\sqrt{\Tilde{D}^{t}_{i}\cdot {D}^{t}_{i} }\right| -\left| \bra{\zeta_t}\ket{\phi^{\prime}_{5}}\right|
\end{equation}
Let us bound the term $\sqrt{\Tilde{D}^{t+1}_{i}\cdot {D}^{t+1}_{i} }$ first.
\begin{equation}
    \begin{split}
        \Tilde{D}^{t+1}_{i}\cdot {D}^{t+1}_{i} &=\sum_{i\in[M]}\sqrt{\frac{\Tilde{D}^{t}_{i}\cdot e^{-\beta^{\prime}_{j,t}}}{\kappa\cdot Z^{\prime}_{t}}\cdot\frac{\Tilde{D}^{t}_{i}\cdot e^{-\beta^{\prime}_{j,t}}}{Z_t}}\\
        &=\sqrt{\frac{Z_t}{\kappa\cdot Z^{\prime}_{t}}}\sum_{i\in[M]}\frac{\Tilde{D}^{t}_{i}\cdot e^{-\beta^{\prime}_{j,t}}}{Z_{t}}=\sqrt{\frac{Z_t}{\kappa\cdot Z^{\prime}_{t}}}=\frac{1}{\sqrt{\kappa}}\cdot\sqrt{\frac{Z_t}{Z^{\prime}_{t}}}\\
        &=\frac{1}{\sqrt{\kappa}}\cdot\sum_{j=1}^{C}\sqrt{\frac{W^{j,t}_{+}e^{-\beta^{\prime}_{j,t}}+W^{j,t}_{-}e^{\beta^{\prime}_{j,t}}}{2\sqrt{\Tilde{W}^{j,t}_{+}\cdot\Tilde{W}^{j,t}_{-}}}}\\
        &=\frac{1}{\sqrt{2\kappa}}\cdot
        \sum_{j=1}^{C}\sqrt{
            \frac{W^{j,t}_{+}e^{-\beta^{\prime}_{j,t}}}
            {\sqrt{\Tilde{W}^{j,t}_{+}\cdot\Tilde{W}^{j,t}_{-}}}
            +\frac{W^{j,t}_{-}e^{\beta^{\prime}_{j,t}}}
            {\sqrt{\Tilde{W}^{j,t}_{+}\cdot\Tilde{W}^{j,t}_{-}}}}\\
        &=\frac{1}{\sqrt{2\kappa}}\cdot
        \sum_{j=1}^{C}\sqrt{
            \frac{W^{j,t}_{+}}{\sqrt{\Tilde{W}^{j,t}_{+}\cdot\Tilde{W}^{j,t}_{-}}}
            \sqrt{\frac{\Tilde{W}^{j,t}_{-}}{\Tilde{W}^{j,t}_{+}}}
            +\frac{W^{j,t}_{-}}
            {\sqrt{\Tilde{W}^{j,t}_{+}\cdot\Tilde{W}^{j,t}_{-}}}\sqrt{\frac{\Tilde{W}^{j,t}_{+}}{\Tilde{W}^{j,t}_{-}}}}\\
        &=\frac{1}{\sqrt{2\kappa}}\cdot
        \sum_{j=1}^{C}\sqrt{\frac{W^{j,t}_{+}}{\Tilde{W}^{j,t}_{+}}
        +\frac{W^{j,t}_{-}}{\Tilde{W}^{j,t}_{-}}}\\
        &\geq \frac{1}{\sqrt{2\kappa}}\cdot
        \sum_{j=1}^{C}\sqrt{\frac{2}{1+\varepsilon}}=\sqrt{\frac{C}{\kappa(1+\varepsilon)}}
    \end{split}
\end{equation}
We also know that $\left| \bra{\zeta_t}\ket{\phi^{\prime}_{5}}\right|\leq\Vert\zeta_t\Vert\leq1-(1-\frac{4\varepsilon}{1+\epsilon})=\frac{4\varepsilon}{1+\epsilon}$. Therefore we have
\begin{equation}
    \left|\bra{\phi_5}\ket{\phi_5^{\prime}}\right|\geq \sqrt{\frac{C}{\kappa(1+\varepsilon)}}-\frac{4\varepsilon}{1+\epsilon}
\end{equation}
Substituting $\kappa=\frac{C}{1-\varepsilon}\sqrt{\frac{1+\varepsilon}{1-\varepsilon}}$ in the above equation, we have
\begin{equation}
    \begin{split}
        \left|\bra{\phi_5}\ket{\phi_5^{\prime}}\right|&\geq \sqrt{\frac{C\cdot(1-\varepsilon)}{C\cdot(1+\varepsilon)}\frac{1-\varepsilon}{1+\varepsilon}}-\frac{4\varepsilon}{1+\epsilon}\\
        &=\left(\frac{1-\varepsilon}{1+\varepsilon}\right)^{3/4}-\frac{4\varepsilon}{1+\epsilon}=\left(1-\frac{2\varepsilon}{1+\varepsilon}\right)^{3/4}-\frac{4\varepsilon}{1+\epsilon}
    \end{split}
\end{equation}
Since $(1-x)^{t}\geq 1-xt, \forall x\leq 1, t>0$, we have
\begin{equation}
        \left|\bra{\phi_5}\ket{\phi_5^{\prime}}\right|\geq 1-\frac{3\varepsilon}{2(1+\varepsilon)}-\frac{4\varepsilon}{1+\epsilon}
        =1-\frac{11\varepsilon}{2(1+\varepsilon)}
\end{equation}
Plugging this back into \eqref{eq:fidelity-eqn}, we get
\begin{equation}
    \begin{split}
        |p-q|\leq2\sqrt{1-\left(1-\frac{11\varepsilon}{2(1+\varepsilon)}\right)^{2Q}}
        &\leq2\sqrt{\frac{11Q\varepsilon}{1+\varepsilon}}<8\sqrt{Q\varepsilon}
    \end{split}
\end{equation}
We now set $\varepsilon=\frac{1}{QT^2}$ which gives us $q=O(1-\frac{1}{T})$ if $p=O(1-\frac{1}{T})$.

\subsection[Proof of margin-bound claim]{Proof of \cref{clm:marginbound} (margin estimation)}\label{subsec:proof_marginbound}
We know that
\begin{equation}\label{eq:relative_error_wjb}
    \left|W^{j,t}_{b}-\Tilde{W}^{j,t}_{b}\right|\leq \varepsilon\cdot W^{j,t}_{b}
\end{equation}
Also, recall that the actual margin given in \cref{alg:RealBoost} is $\beta_{j,t}=\frac{1}{2}\ln{\left(\frac{{W}^{j,t}_{+}}{{W}^{j,t}_{-}}\right)}$ and the estimated margin in \cref{alg:QRealBoost} is  $\beta^{\prime}_{j,t}=\frac{1}{2}\ln{\left(\frac{\Tilde{W}^{j,t}_{+}}{\Tilde{W}^{j,t}_{-}}\right)}$. We upper bound the difference in margins as follows:
\begin{equation}
\label{eq:uppermarginbound}
    \begin{split}
        \beta^{\prime}_{j,t}-\beta_{j,t}&=\frac{1}{2}\left[\ln{\frac{\Tilde{W}^{j,t}_{+}}{\Tilde{W}^{j,t}_{-}}}-\ln{\frac{W^{j,t}_{+}}{W^{j,t}_{-}}}\right] =\frac{1}{2}\left[\ln{\frac{\Tilde{W}^{j,t}_{+}}{W^{j,t}_{+}}}-\ln{\frac{\Tilde{W}^{j,t}_{-}}{{W}^{j,t}_{-}}}\right]\\
        &\leq\frac{1}{2}\left[\ln{\left(1+\varepsilon\right)}-\ln{\left(1-\varepsilon\right)}\right]\\
        &=\frac{1}{2}\ln{\left(\frac{1+\varepsilon}{1-\varepsilon}\right)}\\
    \end{split}
\end{equation}
Similarly, we obtain the lower bound as
\begin{equation}
\label{eq:lowermarginbound}
    \beta^{\prime}_{j,t}-\beta_{j,t}\geq\frac{1}{2}\ln{\left(\frac{1-\varepsilon}{1+\varepsilon}\right)}
\end{equation}
Combining \cref{eq:uppermarginbound} and \cref{eq:lowermarginbound} we get
\begin{equation}
    \left|\beta^{\prime}_{j,t}-\beta_{j,t}\right|\leq \frac{1}{2}\ln{\left(\frac{1+\varepsilon}{1-\varepsilon}\right)}
\end{equation}

\subsection[Proof of normalization const-bound claim]{Proof of \cref{clm:normconstbound} (normalization constant is bounded)}\label{subsec:proof_normconstbound}
The normalization constant in \cref{alg:RealBoost} is calculated as $Z_t=2\sum_{j=1}^{C}\sqrt{{W}^{j,t}_{+}\cdot {W}^{j,t}_{-}}$. In \cref{alg:QRealBoost}, we substitute the weights with out estimated weights to obtain the quantity $Z^{\prime}_{t} = 2\sum_{j=1}^{C}\sqrt{\Tilde{W}^{j,t}_{+}\cdot \Tilde{W}^{j,t}_{-}}$. Using \eqref{eq:relative_error_wjb}, we upper bound the difference between the quantities as
\begin{equation}
\label{eq:uppernormbound}
    \begin{split}
        Z^{\prime}_{t} &= 2\sum_{j=1}^{C}\sqrt{\Tilde{W}^{j,t}_{+}\cdot \Tilde{W}^{j,t}_{-}}\\
        &\leq 2\sum_{j=1}^{C}\sqrt{{W}^{j,t}_{+}\left(1+\varepsilon\right)\cdot {W}^{j,t}_{-}\left(1+\varepsilon\right)}\\
        &=2\left(1+\varepsilon\right)\sum_{j=1}^{C}\sqrt{{W}^{j,t}_{+}\cdot {W}^{j,t}_{-}}\\
        &=Z_t\left(1+\varepsilon\right)
    \end{split}
\end{equation}
Similarly, the lower bound is obtained as
\begin{equation}
\label{eq:lowernormbound}  
    \begin{split}
        Z^{\prime}_{t}&= 2\sum_{j=1}^{C}\sqrt{\Tilde{W}^{j,t}_{+}\cdot \Tilde{W}^{j,t}_{-}}\\
        &\geq 2\sum_{j=1}^{C}\sqrt{{W}^{j,t}_{+}\left(1-\varepsilon\right)\cdot {W}^{j,t}_{-}\left(1-\varepsilon\right)}\\
        &=2\left(1-\varepsilon\right)\sum_{j=1}^{C}\sqrt{{W}^{j,t}_{+}\cdot {W}^{j,t}_{-}}\\
        &=Z_t\left(1-\varepsilon\right)
    \end{split}
\end{equation}
Combining \cref{eq:uppernormbound} and \cref{eq:lowernormbound} we obtain 
\begin{equation}
    \left|Z^{\prime}_{t}- Z_t\right|\leq \varepsilon\cdot Z_t
\end{equation}

\subsection[Proof of zero training error]{Proof of \cref{clm:zerotrainingerror} (final hypothesis has zero training error)}\label{subsec:proof_zerotrainingerror}
For this proof, we follow the framework followed by Freund and Shapire in their book on boosting~\cite{10.5555/2207821}. From \cref{eq:NewUpdateRule} we have
\begin{equation}
    \Tilde{D}^{T+1}_{i} = \frac{\Tilde{D}^{1}_{i}}{\prod_{t=1}^{T}\kappa\cdot Z^{\prime}_{t}}\cdot \mathrm{exp}\left(-y_i\sum_{t=1}^{T}\beta^{\prime}_{j,t}\right)
\end{equation}
Let $x\sim D$ and $\ket{x,0}\xrightarrow[]{O_{h_t}}\ket{x,j^t}$ for all $t\in\{1,\ldots,T\}$ and $F(x)=\sum_{t=1}^T \beta^{\prime}_{j,t}$. Here, note that $H(x)=\mathrm{sign}(F(x))$ according to \cref{eq:qrealboostcombinedclassifier}. Since error means the hypothesis gives a different output than the label, we have, 
\begin{equation}
    \begin{split}
        H(x)\neq y &\implies y\cdot F(x)\leq 0\\
        &\implies \mathrm{exp}(-yF(x))\geq 1\\
        &\implies \mathrm{exp}(-yF(x))\geq \mathbb{I}[H(x)\neq y]
    \end{split}
\end{equation}
The last inequality follows from the fact that $\mathbb{I}[H(x)\neq c(x)]\in\{0,1\}$. Now if we try to upper bound the training error, we have
\begin{equation}
    \begin{split}
        \underset{x\sim \Tilde{D}^{1}}{\mathrm{Pr}}\left[H(x)\neq y\right]&=\sum_{i=1}^{M}\Tilde{D}^{1}_{i}\cdot\mathbb{I}[H(x)\neq y]      \leq\sum_{i=1}^{M}\Tilde{D}^{1}_{i}\cdot\mathrm{exp}(-y_i F(x_i))\\
        &=\sum_{i=1}^{M}\Tilde{D}^{1}_{i}\cdot\mathrm{exp}\left(-y_i\sum_{t=1}^{T}\beta^{\prime}_{j,t}\right)\\
        &=\sum_{i=1}^{M}\Tilde{D}^{T+1}_{i}\prod_{t=1}^{T}\kappa\cdot Z^{\prime}_{t}\\
        &\leq\prod_{t=1}^{T}\kappa\cdot Z^{\prime}_{t}
    \end{split}
\end{equation}
The last inequality is due to the fact that $\sum_{i\in[M]} \Tilde{D}^{t}_{i}\in\left[1-\frac{4\varepsilon}{1+\varepsilon},1\right]$ (as given in \cref{clm:bound_on_subnormalized_sum}). Now from \cref{clm:normconstbound}, we know that $Z^{\prime}_{t}\leq Z_t(1+\varepsilon)$. This means
\begin{equation}
    \begin{split}
         \underset{x\sim \Tilde{D}^{1}}{\mathrm{Pr}}\left[H(x)\neq y\right]&\leq\prod_{t=1}^{T}\kappa\cdot Z_t(1+\varepsilon)         =\kappa^T\left(1+\varepsilon\right)^{T}\prod_{t=1}^{T} Z_t
    \end{split}
\end{equation}
Substituting $\kappa=\frac{C}{\left(1-\varepsilon\right)}\sqrt{\frac{1+\varepsilon}{1-\varepsilon}}$, and the fact that $\varepsilon=O(1/QT^2)$ we have
\begin{equation}
    \begin{split}
        \underset{x\sim \Tilde{D}^{1}}{\mathrm{Pr}}\left[H(x)\neq y\right]&\leq C^{T}\left(\frac{1+1/T^2}{1-1/T^2}\right)^T\prod_{t=1}^{T} Z_t\\
        &\leq C^{T}\left(\frac{T^2+1}{T^2-1}\right)^T\prod_{t=1}^{T} Z_t\\
        &\leq C^{T}\left(1+\frac{2}{T^2-1}\right)^T\prod_{t=1}^{T} Z_t\\
        & \leq C^{T}\mathrm{exp}\left(\frac{2T}{T^2-1}\right)\prod_{t=1}^{T} Z_t\\
    \end{split}
\end{equation}
For sufficiently large $T$, we have
\begin{equation}
    \begin{split}
        \underset{x\sim \Tilde{D}^{1}}{\mathrm{Pr}}\left[H(x)\neq y\right]
        & \leq C^{T}e^{2/T}\prod_{t=1}^{T} Z_t\\
        & \leq C^{T}e^{2/T}\prod_{t=1}^{T} \sqrt{1-4\gamma_t^2}\\
        &\leq C^{T}\mathrm{exp}\left(\frac{2}{T}-2\sum_{t=1}^T\gamma_t^2\right)\\
        & \leq C^{T}\mathrm{exp}\left(\frac{2}{T}-2\gamma^2 T\right)\\
        &\leq C^{T}\mathrm{exp}\left(-2\gamma^2 T+\frac{2}{T}\right)\\
    \end{split}
\end{equation}
The second inequality follows from \cref{thm:trainerr_bound}. We follow the classical analysis from there on to its logical conclusion. For $T=O(\log M/ \gamma^2)$ and a large constant in $O(.)$ \footnote{such that the outer constant is taken care of as well}, we obtain
\begin{equation}
    \underset{x\sim \Tilde{D}^{1}}{\mathrm{Pr}}\left[H(x)\neq y\right]<\frac{1}{M}
\end{equation}
We recall the fact that $\Tilde{D}^1$ is the uniform distribution, which implies that we have zero training error.

\subsection{Proof of Query Complexity}\label{sec:querycomp}
We now analyze the query complexity of our algorithm. The query complexity, as in previous works \cite{Arunachalam2020,deWolf2020}, considers the number of queries to the hypothesis oracles $\left\{O_{h_1},\ldots,O_{h_T}\right\}$ made by \cref{alg:QRealBoost}. We now start calculating the query complexity by considering the queries made in the $t$\textsuperscript{th} iteration.

We require $t-1$ queries to the oracles $O_{h_1},O_{h_2},\ldots,O_{h_{t-1}}$ for each copy of $\ket{\psi}_0$ and $\ket{\phi}_0$. Using \cref{thm:AmplitudeAmplification}, we see that our amplitude amplification algorithm uses an expected $\Theta(p^{\prime}\log{T}/p)$ calls to the unitaries $U_{0\xrightarrow[]{}3}$ and $U^{-1}_{0\xrightarrow[]{}3}$, to obtain $\ket{\phi_4}$ with a high probability (as discussed in \cref{foot:note1}). We observe from $\ket{\phi_3}$ and $\ket{\phi_4}$ that
\begin{equation*}
    p=\sum_{i\in[M]}\sqrt{\Tilde{D}^{t}_{i}/M};\;\;\;\; p^{\prime}=\sum_{i\in[M]}\sqrt{\Tilde{D}^{t}_{i}}
\end{equation*}
Hence, the Amplitude Amplification step to obtain $\ket{\phi_4}$ requires $O(\sqrt{M}\log{T}(t-1))$ queries to the oracles for each copy of $\ket{\phi_3}$. The uncompute step to obtain $\ket{\phi_5}$ requires a further $t-1$ queries to the oracles $O_{h_1},O_{h_2},\ldots,O_{h_{t-1}}$ for each copy of $\ket{\phi_4}$.

For estimating the partition weights with high probability (as discussed in \cref{foot:note1}) we make an expected $\Tilde{O}\left(\sqrt{M}QT^{2}\log{T}\cdot t\right)$ queries to $O_{h_1},O_{h_2},\ldots,O_{h_t}$. We obtain this by plugging in $p=O(1/M)$, $\varepsilon=O(\frac{1}{QT^2})$, and $k=\log{T}$ in \cref{thm:AmplitudeEstimation}. Hence, the total query complexity is
\begin{equation}
    \begin{split}
        &\sum_{t=1}^{T}\left(\underbrace{O(\sqrt{M}Q\log{T}\cdot(t-1))}_{\text{Amplitude Amplification}}+\underbrace{O((Q+C)\log{T}\cdot(t-1))}_{\text{weight updates and uncomputing}}+\underbrace{\Tilde{O}\left(\sqrt{M}CQT^{2}\log{T}\cdot t\right)}_{\text{Amplitude Estimation}}\right)\\
        &=O(\sqrt{M}QT^{2}\log{T})+O((Q+C)T^{2}\log{T})+\Tilde{O}\left(\sqrt{M}CQT^{4}\log{T}\right)\\
        &=\Tilde{O}(\sqrt{M}CQT^{4})=O\left(\frac{\sqrt{M}CQ}{\gamma^{8}}\right)
    \end{split}
\end{equation}
The last equality follows from \cref{thm:trainerr_bound} by setting $T=O(\log{M}/\gamma^{2})$. From \cref{corr:generalizationsamples}, and by setting the parameter $\eta=0.1$, we get the query complexity as $O\left(\frac{\sqrt{d_{\mathcal{H}}}\cdot C\cdot Q}{\gamma^{9}}\right)$.

\subsection{Proof of Time Complexity}\label{sec:timecomp}
We now discuss the time complexity of \cref{alg:QRealBoost}. As discussed in \cref{sec:quantumpaclearning}, we can assume a QRAM to prepare the uniform superposition $\frac{1}{\sqrt{M}}\sum_{i\in [M]}\ket{x_i, y_i, D^{1}_{i}}$ using $O(n\log M)$ gates. Hence the time complexity for preparing the state $\ket{\phi_0}^{\otimes{Q}}\otimes\ket{\psi_0}^{\otimes{2C}}$ is $O(n(Q+C))$. The step from $\ket{\phi_0}$ to $\ket{\phi_1}$ and $\ket{\psi_0}$ to $\ket{\psi_1}$ requires $t-1$ queries each, which can be performed in time $O((Q+C)(t-1))$. Next we perform the distribution update which is an arithmetic operation, using the unitary $U_D$ with the $\ket{j^1_i,\ldots,j^{t-1}_i}$ register as control. This step requires time $O(n^2(Q+C)(t-1))$.

We perform amplitude amplification to obtain the state $\ket{\phi_4}$. This requires $O(\sqrt{M}(t-1)\log T)$ applications of $U^{}_{0\xrightarrow[]{}3}$ and $U^{-1}_{0\xrightarrow[]{}3}$ as discussed in the previous section. The total time taken is therefore $O(n^2\sqrt{M}Q(t-1)\log{T})$. The time taken by our weak learner to output $O_{h_t}$ is $O(n^2 Q)$.

The arithmetic operations to update state $\ket{\psi_3}_{(k,b)}$ to $\ket{\psi_3}_{(k,b)}$ and perform controlled rotation use $O(n)$ gates. Finally we make $\Tilde{O}\left(\sqrt{M}CQT^2\log{T}\right)$ queries for the amplitude estimation part, and each query requires time $O(n^2 t)$. Therefore our final time complexity is
\begin{equation}
\label{eq:timecomplexity}
    \begin{split}
        &\sum_{t=1}^{T}\left(
        \underbrace{O\left(n^2\sqrt{M}Q(t-1)\log{T}\right)}_{\text{Amplitude Amplification}}+
        \underbrace{\Tilde{O}\left(n^2\sqrt{M}CQT^{2}\log{T}\cdot t\right)}_{\text{Amplitude Estimation}}+
        \underbrace{O\left(n^2(Q+C)(t-1)\right)}_{\text{other operations}}\right)\\
        &=\Tilde{O}\left(n^2\sqrt{M}CQT^{4}\log{T}\right)= O\left(\frac{n^2\sqrt{d_{\mathcal{H}}} C Q}{\gamma^{9}}\right)
    \end{split}
\end{equation}
\subsection{Explanation of Laplace Correction}\label{sec:laplace}
Let $V^{k,t}_{b}=\Tilde{W}^{k,t}_{b}\cdot M$. We update the values of $\Tilde{W}^{k,t}$ to $\frac{V^{k,t}_{b}+1}{M+2C}$. We also note that
\begin{equation}
\label{eq:margin_est}
    \beta^{\prime}_{j,t}=\frac{1}{2}\ln{\left(\frac{\Tilde{W}^{j,t}_{+}}{\Tilde{W}^{j,t}_{-}}\right)}\;\;\;\;\;\;\;\forall j\in\{1,\ldots,C\}
\end{equation}
Let us look at the corner cases now. If there exists a partition where $W^{k,t}_{b}=0$ or very small, then we now have $\Tilde{W}^{k,t}_{b}\sim \frac{1}{M+2C}\sim \frac{1}{M}$, essentially resetting the weight. On the other hand, consider a partition where $V^{k,t}_{b}\sim M$. This implies that $V^{k,t}_{-{b}}\sim 0$. By \cref{eq:margin_est}, this would give us unbounded margins $\beta^{\prime}_{k,t}=\frac{1}{2}\ln{\left(\frac{V^{k,t}_{b}}{V^{k,t}_{-b}}\right)\sim\infty}$. Now, due to the smoothing, the confidence for this domain partition will still be large but bounded above by $O(\log{M})$. 

\section{Boosting}\label{sec:adaboost}
Consider the following problem. Let us have query access to a learning algorithm $A$ that is ``weak'', i.e. $A$ performs slightly better than random guessing with respect to some unknown target concept class $\mathcal{C}$. Is it possible for us to obtain a ``strong'' learner that has small empirical and generalization error with respect to $C$, using just oracular access to $A$? 

In \cite{Schapire1990}, Schapire showed us that under the PAC learning model, the task of producing strong learners from weak learners is not only possible but that the two notions of learning are inherently equivalent.
\begin{theorem}[Equivalence of Weak and Strong learning \cite{Schapire1990}]\label{thm:equivalenceoflearning}
An unknown concept class $\mathcal{C}=\bigcup_{n\geq 1}C_n$ is efficiently weakly PAC learnable if and only if $C$ is efficiently strongly PAC learnable.
\end{theorem}

Subsequently, researchers started proposing boosting algorithms (\cite{freund1995boosting,Schapire1990}) (among other types of ensemble learning algorithms) to achieve this task. These algorithms are known as boosting algorithms since they somehow ``boost'' the weak learner to produce a strong learner. These early efforts ultimately culminated in the formation of the {adaptive boosting} or \textbf{AdaBoost} algorithm \cite{Freund1997} (described in \cref{alg:AdaBoost}). A small point to note here is that the definitions of weak and strong learning generalize to the quantum setting simply by considering that the learner is quantum.

% \subsection{Boosting with Binary Weak Predictions}
\begin{algorithm}[t]
\caption{The AdaBoost Algorithm}
\begin{algorithmic}[1]
\label{alg:AdaBoost}
\State\multiline{
\textbf{Input:} Classical weak learner $A$, and $M$ Training Samples $\{(x_1,y_1),(x_2,y_2),\ldots,(x_M,y_M)\}$}; $x_i\sim \mathcal{X},y_i\in\{-1,+1\}$
\State\multiline{
\textbf{Initialize:}  Set ${D^{1}_{i}}=\frac{1}{M}\;\;\forall i\in [M]$ }
\For{$t=1$ to $T$} 
    \State\multiline{ Train $A$ using the distribution $D^{t}$ to obtain the hypothesis $h_t:\mathcal{X}\xrightarrow[]{}\{-1,+1\}$.}
    \State\multiline{ Compute the weighted error $\epsilon_t$ and the margin $\alpha_t$ for this iteration as follows
    \begin{equation}
        \epsilon_t=\sum_{i\in[M]}{D^{t}_{i}[h_t(x_i)\neq y_i]};\;\;\;\alpha_t=\frac{1}{2}\ln{\left(\frac{1-\epsilon_t}{\epsilon_t}\right)}
    \end{equation}}
    \State\multiline{ Set $Z_t=\sum_{i\in[M]}{D^{t}_i}\cdot\mathrm{exp}\left(-\alpha_t y_i h_t(x_i)\right)$.}
  \State\multiline{ Perform the distribution update $\forall i\in\{1,\ldots,M\}$ as follows
    \begin{equation}\label{eq:AdaBoostUpdateRule}
        {D^{t+1}_i}=\frac{{D^{t}_i}\cdot\mathrm{exp}\left(-\alpha_t y_i h_t(x_i)\right)}{Z_{t}}
    \end{equation}}
\EndFor
\textbf{Output:} Hypothesis $H(x)$ where $x\sim \mathcal{X}$.
\begin{equation}
\label{eq:AdaBoostHypo}
    H(x)=\mathrm{sign}\left(\sum_{t=1}^{T}\alpha_t h_t(x_i)\right)
\end{equation}
% \EndProcedure
\end{algorithmic}
\end{algorithm}

\subsection{AdaBoost and some generalizations}
\label{sec:adaboost_gen}
The AdaBoost algorithm takes two inputs - a classical $\gamma$-weak learner $A$, and $M$ training samples. At the $t$\textsuperscript{th} step, we define a new distribution $D^{t}$ over the training samples based on $D^{t-1}$. We then use $A$ to produce a new binary-valued hypothesis with respect to $D^{t}$. We compute the weighted error $\epsilon_t$ and the confidence of the hypothesis $\alpha_t$ for the $t$\textsuperscript{th} step. Using these quantities we define the distribution update rule \cref{eq:AdaBoostUpdateRule} for the next iteration. After at least $T\geq O({\log M}/{\gamma^2})$ iterations we produce the hypothesis $H$ as given in \cref{eq:AdaBoostHypo}. Freund and Schapire showed that $H$ has zero training error and very small generalization error given the number of training samples is sufficiently large.

In AdaBoost for binary classification, we had access to $M$ training samples in $S$ which were distributed according to some unknown distribution $D$ over the domain space $\mathcal{X}$. Given $S$ as input, our weak learner $A$ output a hypothesis $h:\mathcal{X}\xrightarrow[]{}\{-1,+1\}$. Consider the generalized version of the AdaBoost algorithmic framework in which the weak learner outputs real valued hypotheses $h^{\prime}:\mathcal{X}\xrightarrow[]{}\mathbb{R}$. Here $\mathrm{sign}[h^{\prime}(x)]$ is our required prediction, and the quantity $|h^{\prime}(x)|$ gives us the ``confidence'' for the prediction. 

Alternatively, the quantity $|h^{\prime}(x)|$ tells us how confident our learner is while making the prediction $\mathrm{sign}[h^{\prime}(x)]$. This is analogous to the formal notion of margins which is well known in learning theory. Larger margins on training data directly lead us to better bounds on the generalization error \cite{10.5555/2207821}. In fact, boosting algorithms in the AdaBoost framework tend to keep becoming more confident with their predictions leading to a drop in generalization error even after training error converges to zero. 

In \cite{Schapire1999} it was shown that the generalized model with real-valued hypotheses still adheres to the bound given in \cref{thm:trainerr_bound} if $h^{\prime}:\mathcal{X}\xrightarrow[]{}[-1,+1]$. In fact, so far both models of weak learners use \cref{thm:trainerr_bound} to focus on weak learners such that their hypotheses focus on greedily minimizing the normalization constant $Z_t$ (refer \cref{alg:AdaBoost}) at each iteration in order to bound the training error. We can therefore consider folding the margin and the hypothesis into one quantity in order to simplify the calculation of $Z_t$ as in the case of the generalized AdaBoost where $-y_{i}h_{t}^{\prime}(x_i)$ can replace the term $-\alpha_{t}y_{i}h_{t}^{\prime}(x_i)$. In \cref{sec:BoostinDomainPartition} we explore a different flavour of weak learners introduced in \cite{Friedman2000,Schapire1999} that also focus on this particular simplified criteria.

\begin{algorithm}[H]
\caption{The AdaBoost framework}
\begin{algorithmic}[1]
\label{algo:adaboost_framework}
\State\multiline{
\textbf{Input:} Weak learner $A$ with access to $M$ Training samples $\{(x_1,y_1),(x_2,y_2),\ldots,(x_M,y_M)\}$} where $x_i$ are examples and $y_i$ are the corresponding labels.
\State
\textbf{Initialize:}  Set ${D^{1}_{i}}=\frac{1}{M}\;\;\forall i\in [M]$ 
\For{$t=1$ to $T$} 
    \State Train $A$ using the distribution $D^{t}$ to obtain the hypothesis $h_t$.
    \State Compute the weighted misclassification error over all samples and the confidence of the hypothesis for this iteration by comparing the predicted value and the label for each sample.
    \State  Update the distribution to $D^{t+1}$ using the computed errors and confidences.
\EndFor
\State \textbf{Output:} Hypothesis $H(x)$ which combines the individual hypothesis $h_t, \forall t\geq 1$ according to their computed confidences.
% \EndProcedure
\end{algorithmic}
\end{algorithm}

\begin{algorithm}[h]
\caption{The RealBoost Algorithm}
\begin{algorithmic}[1]
\small
\label{alg:RealBoost}
\State\multiline{
\textbf{Input:} Classical weak learner $A$, and $M$ training samples $\{(x_1,y_1),(x_2,y_2),\ldots,(x_M,y_M)\}$}; $x_i\sim \mathcal{X},y_i\in\{-1,+1\}$
\State\multiline{
\textbf{Initialize:}  Set ${D^{1}_{i}}=\frac{1}{M}\;\;\forall i\in [M]$ }
\For{$t=1$ to $T$} 
    \State Train $A$ using the distribution $D^{t}$ to obtain a partitioning $\mathcal{X}^t=\{\mathcal{X}^t_1, \ldots, \mathcal{X}^t_C\}$ of  $\mathcal{X}$ % $j^t_i\in\{1,2,\ldots,C\}$ for the sample $x_i$, for all $i \in [M]$.}
    \For{$k=1$ to $C$}\Comment{We iterate over every partition.}
        \For{$b\in\{-1,+1\}$}\Comment{We iterate over every label.}
            \State\multiline{ Calculate the partition weight $W^{k,t}_{b}$ as
            \begin{equation}\label{eq:realboostpartition}
                W^{k,t}_{b}=\sum_{i:x_i\in \mathcal{X}^t_k \textbf{ and }y_i=b}D^{t}_i
            \end{equation}}
        \EndFor
        % \State\multiline{}
        % \begin{equation}
        %     \beta_{j,t}=\frac{1}{2}\ln{\left(\frac{W^{k,t}_{+}}{W^{k,t}_{-}}\right)}
        % \end{equation}}
    \EndFor
    \State Set $Z_t=2\sum_{j=1}^{C}{\sqrt{W^{j,t}_{+}\cdot W^{j,t}_{-}}}$
    \State Compute the margins $\beta_{j,t}=\frac{1}{2}\ln{\left(\frac{W^{j,t}_{+}}{W^{j,t}_{-}}\right)},\forall j\in\{1,\ldots,C\}$.
   \State\multiline{ Perform the distribution update $\forall i\in\{1,\ldots,M\}$ as follows
    \begin{equation}\label{eq:RealBoostUpdateRule}
        {D^{t+1}_i}=\frac{{D^{t}_i}\cdot\mathrm{exp}\left(-\beta_{j,t}\cdot y_i\right)}{Z_{t}}
    \end{equation}
    where $x_i\in \mathcal{X}^t_j$.}
\EndFor
\State{\textbf{Output:} Hypothesis $H(\cdot)$ which is defined as
\begin{equation}
    H(x)=\mathrm{sign}\left(\sum_{t=1}^{T}\beta_{j,t}\right) \mbox{~ where $j$ indicates the domain partition $\mathcal{X}^t_j \in \mathcal{X}^t$ containing $x$.}
\end{equation}}
% \EndProcedure
\end{algorithmic}
\end{algorithm}

\subsection{Error Bounds and Sample Complexity}
In this section, we state a few theorems and definitions related to training and generalization error bounds as well as sample complexity of the algorithms that follow the AdaBoost framework, for example, RealBoost (\cref{alg:RealBoost}).
%This includes both \cref{alg:AdaBoost} and \cref{alg:RealBoost}.

\begin{theorem}[Upper Bound on Training Error \cite{10.5555/2207821}]
\label{thm:trainerr_bound}
Let $A$ be a $\gamma$-weak learner. Let $\gamma_t=\frac{1}{2}-\epsilon_t$, where $\epsilon_t$ is misclassification error at every iteration of \cref{alg:RealBoost}. Let $D_1$ be an arbitrary initial distribution over the training set. Then the weighted training error of the combined classifier $H$ with respect to $D_1$ is bounded as
\begin{equation}
    \Hat{\mathrm{err}}(H)\leq\prod_{t=1}^{T}Z_t\leq\prod_{t=1}^{T}\sqrt{1-4\gamma_t^2}\leq\mathrm{exp}\left(-2\sum_{t=1}^T\gamma_t^2\right)
\end{equation}
\end{theorem}

If we look at the first inequality in \cref{thm:trainerr_bound}, we observe that the AdaBoost framework minimizes the training error of the combined hypothesis by greedily minimizing the normalization constant in \cref{alg:RealBoost} at every step. This produces the following corollary that we shall use later.

\begin{corollary}
\label{corr:algorithmconvergence}
Let $A$ be a $\gamma$-weak learner and $D_1$ be the uniform distribution over the training set of $M$ examples. Then the training error of the combined classifier $H$ with respect to $D_1$ goes to $0$ when $T>\frac{\ln{M}}{2\gamma^2}$, where $T$ is the number of iterations of our boosting algorithm.
\end{corollary}

We see that greedily minimizing the training error at every step leads to the algorithm converging exponentially fast in terms of training samples. 
% We now define the Vapnik-Chervonenkis dimension (or more commonly referred to as the VC-dimension) which is an important quantity that is used to bound the generalization error of a hypothesis class $\mathcal{H}$.

% \begin{definition}[VC-dimension]
% Let $C=\bigcup_{n\geq 1}C_n$ be a concept class over the domain $\mathcal{X}_n$. Consider a set of points $\mathcal{S}\subseteq\mathcal{X}_n$. The concept class $\mathcal{C}$ shatters the set of points $\mathcal{S}$ if, for every labelling $\ell:\mathcal{S}\xrightarrow[]{}\{-1,+1\}$ there exists a $c\in C_n$ such that $c(s)=\ell(s),\forall s\in\mathcal{S}$. Alternatively, a hypothesis class $\mathcal{H}$ shatters $\mathcal{S}$ if there exists $h\in\mathcal{H}$ such that $h(s)=\ell(s),\forall s\in\mathcal{S}$. We denote the VC-dimension of $\mathcal{H}$ (respectively $C$) by $d_{\mathcal{H}}$ (respectively $d_C$) as the largest set of points $\mathcal{S}$ that can be shattered by $\mathcal{H}$ (respectively $C$).
% \end{definition}

% We can observe that if $d_{\mathcal{H}}$ is small, then it should be relatively easier to find a good hypothesis $h\in\mathcal{H}$, i.e., a hypothesis that minimizes training error. 
The next theorem gives us bounds on the generalization error in the AdaBoost framework.

\begin{theorem}[Generalization Error Bound \cite{10.5555/2207821}]
\label{thm:generalizationerrorbound}
Let us have a $\gamma$-weak learner $A$ for a concept class $\mathcal{C}$ which produces classifiers $h$ from a space $\mathcal{H}$ which has finite VC-dimension $d_{\mathcal{H}}\geq 1$. If we run \cref{alg:RealBoost} for $T$ rounds on $M$ random samples (sampled from an unknown distribution $D:\{0,1\}^{n}\xrightarrow[]{}[0,1]$ and associated with a concept class $\mathcal{C}= \bigcup_{n\geq1}C_{n}$) such that $m\geq\mathrm{max}\{d_{\mathcal{H}},T\}$, then with high probability (at least $2/3$), the final hypothesis $H:\{0,1\}^{n}\xrightarrow[]{}\{-1,+1\}$ satisfies
\begin{equation}
    \label{eq:generalizationerrorbound}
    \mathrm{err}(H)\leq \Hat{\mathrm{err}}(H)+\Tilde{O}\left(\sqrt{\frac{Td_{\mathcal{H}}}{M}}\right)
\end{equation}
\end{theorem}

From \cref{thm:generalizationerrorbound}, we get the following corollary which helps us lower bound the total number of training samples to obtain a low generalization bound for \cref{alg:RealBoost}.

\begin{corollary}[Sample Complexity for reducing generalization error of AdaBoost]
\label{corr:generalizationsamples}
Let us have a $\gamma$-weak learner $A$ which produces classifiers $h$ from a space $\mathcal{H}$ which has finite VC-dimension $d_{\mathcal{H}}\geq 1$. We sample $M$ random samples from an unknown distribution $D:\{0,1\}^{n}\xrightarrow[]{}[0,1]$ which are associated with a concept class $\mathcal{C}= \bigcup_{n\geq1}C_{n}$ such that $m\geq\mathrm{max}\{d_{\mathcal{H}},T\}$. If we run \cref{alg:RealBoost} for $T$ rounds where $T\geq \Tilde{O}(\log{M}/\gamma^{2})$, then with high probability we get a generalization error of at most $\eta>0$ when
\begin{equation*}
    M\geq \Tilde{O}\left(\frac{d_{\mathcal{H}}}{\gamma^{2}\eta^{2}}\right)
\end{equation*}
\end{corollary}
\end{document}